\pdfoutput=1
\documentclass[prd,preprint,superscriptaddress,preprintnumbers,eqsecnum,showpacs,nofootinbib,nobibnotes]{revtex4-1}
\usepackage{amsfonts,amsmath,amssymb}
\usepackage{graphicx} 

\newcommand{\be}{\begin{equation}}
\newcommand{\bea}{\begin{eqnarray}}
\newcommand{\ee}{\end{equation}}
\newcommand{\eea}{\end{eqnarray}}

\def\s#1{{\scriptscriptstyle #1}}

\def\noeq#1{(\ref{#1})}
\def\1eq#1{Eq.~(\ref{#1})}

\def\2eqs#1#2{Eqs.~(\ref{#1}) and~(\ref{#2})}
\def\3eqs#1#2#3{Eqs.~(\ref{#1}),~(\ref{#2}) and~(\ref{#3})}

\def\fig#1{Fig.~\ref{#1}}

\def\kint{\int_k\!}
\def\diff{{\rm d}}

\def\ie{{\it i.e.}, }
\def\eg{{\it e.g.}, }

\def\n#1{$(#1)$}

\def\g{\widetilde\Gamma^{\mathrm{np}}}

\def\hh{m^2}

\def\cd{\!\cdot\!}

\def\Cgh{\widetilde{C}_{\mathrm{gh}}}
\def\Cgl{\widetilde{C}_{\mathrm{gl}}}

\def\Sgh{\widetilde{S}_{\mathrm{gh}}}
\def\Sgl{\widetilde{S}_{\mathrm{gl}}}

\def\s#1{{\scriptscriptstyle #1}}

\def\fgl{f_{\mathrm{gl}}}
\def\fgh{f_{\mathrm{gh}}}
\def\xiQ{\xi_\s{\mathrm{Q}}}

\def\aBSE{\alpha_s^\s{\mathrm{BSE}}}
\def\aDSE{\alpha_s^\s{\mathrm{SDE}}}

\begin{document}

\title{Evidence of ghost suppression in gluon mass dynamics}

\author{A.~C. Aguilar}
%\email{aguilar@ifi.unicamp.br}
\affiliation{University of Campinas - UNICAMP, 
Institute of Physics ``Gleb Wataghin'',
13083-859 Campinas, SP, Brazil}

\author{D. Binosi}
%\email{binosi@ectstar.eu}
\affiliation{European Centre for Theoretical Studies in Nuclear
Physics and Related Areas (ECT*) and Fondazione Bruno Kessler, \\Villa Tambosi, Strada delle
Tabarelle 286, 
I-38123 Villazzano (TN)  Italy}

\author{C.~T. Figueiredo}
%\email{aguilar@ifi.unicamp.br}
\affiliation{University of Campinas - UNICAMP, 
Institute of Physics ``Gleb Wataghin'',
13083-859 Campinas, SP, Brazil}

\author{J. Papavassiliou}
%\email{Joannis.Papavassiliou@uv.es}
\affiliation{\mbox{Department of Theoretical Physics and IFIC, 
University of Valencia and CSIC},
E-46100, Valencia, Spain}

\begin{abstract}

In  this work  we study  the  impact that  the ghost  sector of  pure
 Yang-Mills theories may  have on the generation of  a dynamical gauge
 boson mass, which  hinges on the appearance of massless  poles in the
 fundamental vertices of the theory, and the subsequent realization of
 the well-known Schwinger mechanism.   The process responsible for the
 formation of  such structures is  itself dynamical in nature,  and is
 governed  by a  set  of Bethe-Salpeter  type  of integral  equations.
 While in previous studies the  presence of massless poles was assumed
 to be  exclusively associated  with the  background-gauge three-gluon
 vertex, in the  present analysis we allow them to  appear also in the
 corresponding ghost-gluon vertex.  The full analysis of the resulting
 Bethe-Salpeter  system reveals  that  the contribution  of the  poles
 associated with  the ghost-gluon vertex are  particularly suppressed,
 their sole  discernible effect  being a  slight modification  in the
 running of  the gluon mass,  for momenta larger  than a few  GeV.  In
 addition,  we   examine  the   behavior  of   the  (background-gauge)
 ghost-gluon  vertex in  the limit  of vanishing  ghost momentum,  and
 derive the corresponding version of Taylor's theorem.
 These considerations, together with a suitable {\it Ansatz}, 
 permit us the full reconstruction of the
 pole sector of the two vertices involved.

\end{abstract}


\maketitle

\section{\label{sec:intro} Introduction}                                                      
The nonperturbative generation of an effective gluon mass 
has attracted particular attention in the last decade,
being identified as one of the fundamental emergent phenomena produced by the 
intricate gauge-sector dynamics of QCD~\cite{Cloet:2013jya,Roberts:2016vyn,Roberts:2016mhh}.
As has been advocated in a series of works~\cite{Aguilar:2006gr,Aguilar:2008xm, Aguilar:2011xe,Ibanez:2012zk,Aguilar:2016vin},
the appearance of such a (momentum-dependent) mass~\cite{Cornwall:1981zr}, $m^2(q^2)$, is inextricably connected with the
infrared finiteness of the gluon propagator,  $\Delta(q^2)$, 
and the ghost dressing function, $F(q^2)$, observed in a variety of
large-volume lattice simulations~\cite{Cucchieri:2007md,Cucchieri:2007rg,Cucchieri:2009zt,Bowman:2007du,Bogolubsky:2009dc,Oliveira:2009eh,Ayala:2012pb,Bicudo:2015rma}.
Even though these paradigm-shifting lattice results have been explained and interpreted within a plethora of diverse theoretical
approaches~\cite{Cornwall:1981zr,Lavelle:1991ve,Halzen:1992vd,Philipsen:2001ip,Szczepaniak:2001rg,Aguilar:2004sw,Aguilar:2006gr,Kondo:2006ih,Braun:2007bx,Epple:2007ut,Aguilar:2008xm,Boucaud:2008ky,Dudal:2008sp,Fischer:2008uz,Aguilar:2009nf,RodriguezQuintero:2010wy,Campagnari:2010wc,Tissier:2010ts,Kondo:2010ts,Pennington:2011xs,Watson:2011kv,Kondo:2011ab,Serreau:2012cg,Strauss:2012dg,Cloet:2013jya,Siringo:2014lva,Binosi:2014aea,Aguilar:2015nqa,Huber:2015ria,Capri:2015ixa,Binosi:2016nme,Glazek:2017rwe,Gao:2017uox}, in the present work we 
employ the formal framework that emerges from the
fusion between the pinch-technique (PT)~\cite{Cornwall:1981zr,Cornwall:1989gv,Pilaftsis:1996fh,Binosi:2002ft,Binosi:2003rr,Binosi:2009qm} with the  background-field method (BFM)~\cite{Abbott:1980hw}, known as ``PT-BFM scheme''~\cite{Aguilar:2006gr,Binosi:2007pi,Binosi:2008qk}. 

The set of  basic ideas underlying the approach put  forth in~\cite{Aguilar:2011xe,Ibanez:2012zk}, and
more recently  in~\cite{Aguilar:2016vin}, may be  summarized as follows.  At  the level of
the Schwinger-Dyson equation  (SDE) that governs the  dynamics of the
gluon propagator within the PT-BFM scheme,
the masslessness of the gluon is  enforced  nonperturbatively
by   means  of  a  special  integral
identity (``seagull'' identity~\cite{Aguilar:2009ke,Aguilar:2016vin}). This identity is 
triggered by the special (Abelian) Slavnov-Taylor identities (STIs) satisfied
by the fundamental vertices appearing in  the diagrammatic expansion of the 
gluon  SDE\footnote{We remind the reader that, within the PT-BFM scheme,
at least one of the two legs entering into the gluon propagator is 
a ``background'' gluon (see next section). All such vertices
are generically denoted by $\widetilde{\Gamma}$, while their conventional
counterparts by ${\Gamma}$.}, enforcing the exact result $\Delta^{-1}(0)=0$. 
The action of the seagull identity  may be  circumvented,
allowing for the possibility $\Delta^{-1}(0) \neq 0$, only  if the  well-known
Schwinger  mechanism~\cite{Schwinger:1962tn,Schwinger:1962tp}  is triggered~\cite{Jackiw:1973tr,Smit:1974je,Eichten:1974et,Poggio:1974qs}. The  activation  of this  latter
mechanism, in  turn, requires  the presence of
longitudinally coupled massless poles, \ie of the generic form $(q^{\mu}/q^2) \widetilde {C}(q,r,p)$, 
in the aforementioned vertices entering in the gluon SDE.

The  origin of  these  poles is dynamical rather than
kinematic, and may  be traced  back to  the
formation   of  tightly   bound  {\it   colored}  excitations;
in fact, within this picture, the terms $\widetilde {C}$
may be identified
with the ``bound-state  wave   functions'' of these excitations. 
The quantities relevant for the generation of a gluon mass and 
the determination of its momentum dependence are the partial derivatives of the
$\widetilde {C}(q,r,p)$ as $q\to 0$,  to be generically denoted by $\widetilde {C}^{\prime}(r^2)$; 
their evolution, in turn, is controlled by a  system  of coupled 
homogeneous linear Bethe-Salpeter equations (BSEs)~\cite{Jackiw:1973tr,Smit:1974je,Eichten:1974et,Poggio:1974qs}. 

Even though, in principle,  all fundamental vertices 
entering into the gluon SDE, \ie the three-gluon, ghost-gluon, and four-gluon vertex, 
may develop such poles, one of the main simplifications
implemented in all previous studies is the assumption that the
dominant effect originates from the three-gluon vertex, and that all contributions from the 
pole parts of the remaining vertices are numerically subleading.
This assumption, in turn, reduces dramatically the level of technical complexity, 
converting the system of coupled BSEs into one single
dynamical equation (in the Landau gauge). 
In the present work we partially relax this basic assumption  
by including massless poles also in the ghost-gluon vertex, $\widetilde{\Gamma}_{\mu}$,
and studying in detail how the results previously obtained are
affected by their presence\footnote{Note however that 
we are still operating 
under the hypothesis that potential effects due to poles
associated with the four-gluon vertex are numerically suppressed.}. 

The analysis necessary for addressing the aforementioned dynamical question is significantly more
complicated than that of~\cite{Aguilar:2011xe,Binosi:2017rwj}, mainly due to the fact that the
pole formation is now governed by a system of two coupled
integral equations. Specifically, the resulting system of BSEs involves
as unknown quantities the derivative of the wave function of the pole
in the three-gluon vertex, $\widetilde{\Gamma}_{\mu\alpha\beta}$,  
to be denoted by $\Cgl^{\prime}(r^2)$, and the corresponding quantity in $\widetilde{\Gamma}_{\mu}$, to
be denoted by $\Cgh^{\prime}(r^2)$.

These two quantities affect the gluon dynamics in rather distinct ways.
To begin with, both $\Cgl^{\prime}(r^2)$ and $\Cgh^{\prime}(r^2)$ enter in the formula that
determines the value of $\Delta^{-1}(0)$ [see \1eq{DSEmass}];
however, their relative contribution can be vastly different,
even if it turned out that $\Cgl^{\prime}(r^2) \simeq \Cgh^{\prime}(r^2)$, because  
they are convoluted with completely different structures.
Moreover, as has been shown first in~\cite{Aguilar:2011xe} and recently revisited in~\cite{Binosi:2017rwj}, 
the running gluon mass, $m^2(q^2)$, is entirely determined from the form of $\Cgl^{\prime}(r^2)$.
Therefore, the way that $\Cgh^{\prime}(r^2)$ could 
affect $m^2(q^2)$ is indirect, depending on the difference 
between the $\Cgl^{\prime}(r^2)$ found from the (single) BSE when 
$\Cgh^{\prime}(r^2)$ is assumed to vanish identically, as was done previously~\cite{Aguilar:2011xe,Binosi:2017rwj}, 
and the $\Cgl^{\prime}(r^2)$ obtained by actually solving the coupled BSE system, as we do here. 

The full analysis of the BSE system 
carried out in the present work reveals that $\Cgh^{\prime}(r^2)$ is considerably 
smaller than $\Cgl^{\prime}(r^2)$. Specifically, 
when all quantities entering into the kernels of the BSE system  
have been renormalized using the momentum subtraction scheme (MOM) at the point $\mu =4.3$ GeV, 
the relative size between the two quantities is approximately
$\Cgh^{\prime}(r^2)/\Cgl^{\prime}(r^2) \simeq 1/5$. As a result, the substitution of
$\Cgl^{\prime}(r^2)$ and $\Cgh^{\prime}(r^2)$ into the corresponding integrals
that determine $\Delta^{-1}(0)$ shows that the effect stemming from $\Cgh^{\prime}(r^2)$
is practically negligible.
This conclusion may be restated 
in terms of the quadratic equation for the strong coupling
$\alpha_s$, introduced in~\cite{Binosi:2017rwj}; specifically, the 
value of $\alpha_s$ that emerges from the combination of the BSE and the SDE  
remains practically unchanged in the presence of the nonvanishing, but rather small, $\Cgh^{\prime}(r^2)$.
The only place where $\Cgh^{\prime}(r^2)$ makes a small but discernible difference
is in the running of $m^2(q^2)$, in the region of momenta more than a few GeV. 
In particular, the deviation from the exact power-law running is controlled by the value of
the exponent $p$, which changes from the value $p=0.1$ when $\Cgh^{\prime}(r^2)$ is neglected~\cite{Binosi:2017rwj} 
to the value $p=0.24$ when $\Cgh^{\prime}(r^2)$ is included. Thus, the overall conclusion of this
work is that the effects of the ghost sector, in the sense described above, do not modify appreciably the 
dynamics responsible for the generation of an effective gluon mass.

In addition to the findings just mentioned, the present study  
addresses certain aspects related to the structure and behaviour of $\widetilde{\Gamma}_{\mu}$,
which are theoretically interesting and novel,
and furnish further insights into the underlying mass-generation mechanism.
Specifically, as is well-known, in the limit of vanishing ghost momentum, 
the form-factors of the conventional ghost-gluon vertex, ${\Gamma}_{\mu}$, satisfy a special exact relation,
known as {\it Taylor's theorem}~\cite{Taylor:1971ff}. In this work we derive the corresponding relation for
$\widetilde{\Gamma}_{\mu}$, using three vastly different approaches. 
The form of Taylor's theorem that emerges is clearly different from the standard case,
involving the ghost dressing function $F(q^2)$ as its new main ingredient. 

Furthermore, the structure of $\widetilde{\Gamma}_{\mu}$ is scrutinized, placing
particular emphasis on the way that the
fundamental (Abelian) STI is realized in the presence of a longitudinally coupled pole term.
In fact, it is shown that through an appropriate rearrangement of its form factors,
consistent with the (newly derived) version of Taylor's theorem,  
the effect of the pole may be reabsorbed in the transverse (automatically conserved)
part of the vertex. The above considerations are not without practical interest, since 
they allow us to fully determine (under some mild assumptions) the entire function  $\Cgh(q,r,p)$ from the knowledge of $\Cgh^{\prime}(r^2)$.

The article is organized as follows.
In Sec.~\ref{sec:framework} we review the basic formalism
employed in this work, with particular emphasis on the
way the massless poles enter into the vertices, and the
special way the corresponding STIs are satisfied in their
presence.
Then, in Sec.~\ref{sec:taylor} we derive the
version of Taylor's theorem applicable to $\widetilde{\Gamma}_{\mu}$,
using three different procedures: \n{i} the STI that $\widetilde{\Gamma}_{\mu}$ satisfies;
\n{ii} the SDE of $\widetilde{\Gamma}_{\mu}$, and \n{iii}
an exact relation
connecting $\widetilde{\Gamma}_{\mu}$ with ${\Gamma}_{\mu}$, known as
``background-quantum identity'' (BQI)~\cite{Binosi:2008qk}.
In Sec.~\ref{sec:polepart} we offer a new perspective on the way that
the STI of $\widetilde{\Gamma}_{\mu}$ is enforced for a nonvanishing $\Cgh(q,r,p)$,
as well as the constraints imposed on it from Taylor's theorem. The upshot of this 
analysis is the demonstration that  one may reinterpret the action of the longitudinally coupled pole  
as a corresponding pole contribution in the transverse part of $\widetilde{\Gamma}_{\mu}$.   
In addition, using the above results, we present a simple {\it Ansatz} for $\Cgh(q,r,p)$, which
allows its full reconstruction, once $\Cgh^{\prime}(r^2)$ has been determined.
In Sec.~\ref{furcon} we derive the BSE system that governs $\Cgl^{\prime}(r^2)$ and $\Cgh^{\prime}(r^2)$.
Then, in Sec.~\ref{numan} we present the numerical analysis, and establish the
subleading nature of the ghost-related contributions.
Finally, in Sec.~\ref{conc} we present our conclusions.

\section{\label{sec:framework} Gluon mass from vertices with  massless poles}                             
For an SU(3) pure Yang-Mills theory (no dynamical quarks) quantized in the Landau gauge, the gluon and ghost propagators have the form (we factor out the trivial color structure $\delta^{ab}$)
\begin{align}
	\Delta_{\mu\nu}(q) &= -i\Delta(q^2)P_{\mu\nu}(q); \qquad P_{\mu\nu}(q) = g_{\mu\nu}-\frac{q_\mu q_\nu}{q^2},\nonumber \\
	D(q^2)&=i\frac{F(q^2)}{q^2}.
\end{align}  
In the formulas above, $\Delta(q^2)$ is related to the scalar form factor of the gluon self-energy  \mbox{$\Pi_{\mu\nu}(q) = P_{\mu\nu}(q) \Pi (q^2)$} through $\Delta^{-1}(q^2) = q^2 + i \Pi (q^2)$, while $F(q^2)$ represents the so-called ghost dressing function; at tree-level $\Delta^{(0)}(q^2)=1/q^2$ and $F^{(0)}(q^2)=1$.

Within the PT-BFM framework that we employ in the ensuing analysis\footnote{Inherent to such framework is the distinction between background ($B$) and quantum ($Q$) gluons, the proliferation of the possible Green’s functions that one may form with them, and the relations they have. In the following, functions involving $B$ fields will carry a tilde.}, the SDE of $\Delta(q^2)$ is expressed in terms of the $QB$ self-energy $\widetilde\Pi_{\mu\nu}(q)$, namely  
%(see~\fig{fig:procedure})
\begin{align}
	\Delta^{-1}(q^2)P_{\mu\nu}(q) &= \frac{q^2P_{\mu\nu}(q)  + i
	%[\overbrace{\a_{\mu\nu}(q) + \b_{\mu\nu} +\c_{\mu\nu}(q) + \d_{\mu\nu}+\e_{\mu\nu}(q)+\f_{\mu\nu}(q)}^{\widetilde{\Pi}_{\mu\nu}(q)}]
	\widetilde{\Pi}_{\mu\nu}(q)}{1 + G(q^2)}, 
\label{glSDE1}
\end{align}
where $G(q^2)$ is the $g_{\mu\nu}$ component of a special two-point function~\cite{Binosi:2002ez}, related to the ghost dressing function through the equation\footnote{This result originates from 
the identity in~\1eq{funrel}, which is valid only in the Landau gauge. Its generalization to linear covariant gauges involves an additional auxiliary function, and can be found in~\cite{Binosi:2013cea}.} \mbox{$1+G(0) = F^{-1}(0)$}, see also~\1eq{funrel} below~\cite{Grassi:2004yq,Aguilar:2009nf}. 

Expressing the gluon SDE in terms of $\widetilde\Pi_{\mu\nu}(q)$ 
rather than $\Pi_{\mu\nu}(q)$ entails the advantage that, 
when contracted from the side of the $B$-gluon, each fully dressed vertex 
satisfies a linear (Abelian-like) Slavnov-Taylor identity (STI). 
In particular, the $BQ^2$ vertex $\widetilde{\Gamma}_{\mu\alpha\beta}$ and the $Bc\bar c$ vertex $\widetilde{\Gamma}_{\mu}$ satisfy (color omitted and all momenta entering)
\begin{align}
	&q^\mu \widetilde{\Gamma}_{\mu\alpha\beta}(q,r,p) = i\Delta_{\alpha\beta}^{-1}(r) - i\Delta_{\alpha\beta}^{-1}(p),\label{AbWI3gl}\\
	&q^\mu \widetilde{\Gamma}_\mu(q,r,p) = iD^{-1}(r^2) - iD^{-1}(p^2),\label{AbWI3gh}
\end{align}
whereas for the $BQ^3$ vertex we have
\begin{align}
    q^\mu \widetilde{\Gamma}^{mnrs}_{\mu\alpha\beta\gamma}(q,r,p,t) &= f^{mse}f^{ern} \Gamma_{\alpha\beta\gamma}(r,p,q+t) + f^{mne}f^{esr}\Gamma_{\beta\gamma\alpha}(p,t,q+r) \nonumber \\
	&+ f^{mre}f^{ens} \Gamma_{\gamma\alpha\beta}(t,r,q+p).
	\label{AbWI4gl}
\end{align}
%The corresponding STI satisfied by the four-gluon vertex is of no interest to the present analysis, and will be omitted.

Recently, it has been shown that if the 
vertices carrying the $B$ leg do not contain massless poles of the type 
$1/q^2$, then the $\Delta(q^2)$ governed by \1eq{glSDE1} remains rigorously  massless~\cite{Aguilar:2016vin}. The demonstration relies on the subtle interplay between the Ward-Takahashi identities (WTIs), satisfied by the vertices as $q\to 0$, and an integral relation known as the ``seagull identity''~\cite{Aguilar:2009ke,Aguilar:2016vin}. In fact, in the absence of massless poles, the Taylor expansion of both sides of~\2eqs{AbWI3gl}{AbWI3gh} generates the corresponding WTIs
\begin{align}
    &\widetilde{\Gamma}_{\mu\alpha\beta}(0,r,-r) = -i \frac{\partial }{\partial r^\mu}\Delta^{-1}_{\alpha\beta}(r),\label{WTI3gl}\\
    &\widetilde{\Gamma}_{\mu}(0,r,-r) = -i \frac{\partial }{\partial r^\mu}D^{-1}(r^2),
    \label{WTI3gh}	
\end{align}
and
\begin{align}
    \widetilde{\Gamma}^{mnrs}_{\mu\alpha\beta\gamma}(0,-r,-p,r+p) &= -\left(f^{mne}f^{esr}\frac{\partial}{\partial r^\mu} + f^{mre}f^{ens}\frac{\partial}{\partial p^\mu}\right)\Gamma_{\alpha\beta\gamma}(-r,-p,r+p).
    \label{WTI4gl}
\end{align}

Using these expressions in evaluating the gluon SDE, yields then\footnote{We define the  dimensional regularization integral measure $\int_{k}\equiv\frac{\mu^{\epsilon}}{(2\pi)^{d}}\!\int\!\mathrm{d}^d k$, with $d=4-\epsilon$ the space-time dimension, and $\mu$  the 't Hooft mass scale.} 
\begin{align}
    \Delta^{-1}(0) & = \underbrace{\int_k\frac{\partial}{\partial k_\mu}{\cal F}_\mu(k) =0}_{\rm seagull \,\, identity},
    \label{seag1}
\end{align}
where
\begin{align}
    {\cal F}_\mu(k)&=k_\mu{\cal F}(k^2);&
    {\cal F}(k^2) = \Delta(k^2) [c_1 + c_2 Y(k^2)]+c_3D(k^2),
\end{align}
with $c_1,c_2,c_3 \neq 0$, and 
\begin{align}
    Y(k^2)&= \frac{1}{(d-1)}\frac{k_\alpha}{k^2}\! \int_\ell\!\Delta^{\alpha\rho}(\ell)\Delta^{\beta\sigma}(\ell+k)\Gamma_{\sigma\rho\beta}(-\ell-k,\ell,k).
    \label{defY}    
\end{align}

This result may be circumvented by relaxing the assumption made when deriving \2eqs{AbWI3gl}{AbWI3gh}, allowing the vertices to contain longitudinally coupled $1/q^2$ poles; their inclusion, 
in turn, triggers the Schwinger mechanism~\cite{Schwinger:1962tn,Schwinger:1962tp}, 
finally enabling the generation of a gauge boson mass~\cite{Jackiw:1973tr,Smit:1974je,Eichten:1974et,Poggio:1974qs}. 

Neglecting effects stemming from poles associated with the four-gluon vertex, the $BQ^2$ and $Bc\bar c$ vertices will then take the form 
\begin{align}
    \widetilde{\Gamma}_{\mu\alpha\beta}(q,r,p) &= \g_{\mu\alpha\beta}(q,r,p) + i \frac{q_\mu}{q^2}\widetilde{C}_{\alpha\beta}(q,r,p),
    \label{GnpGp1}\\ 
    \widetilde{\Gamma}_{\mu}(q,r,p)&=\g_{\mu}(q,r,p)+ i \frac{q_\mu}{q^2}\Cgh(q,r,p), 
    \label{GnpGp2}
\end{align}
where the superscript ``np'' stands for ``no-pole'', whereas $\widetilde{C}_{\alpha\beta}$ and $\Cgh$ represents the bound-state gluon-gluon and gluon-ghost wave functions, respectively~\cite{Jackiw:1973tr,Eichten:1974et,Poggio:1974qs}.

Next, in order to preserve the BRST symmetry of the theory, 
we demand that all STIs maintain their exact form in the presence of these poles; therefore, \2eqs{AbWI3gl}{AbWI3gh} will now read  
\begin{align}
	& q^\mu \g_{\mu\alpha\beta}(q,r,p) + \widetilde{C}_{\alpha\beta}(q,r,p) = i\Delta_{\alpha\beta}^{-1}(r) - i\Delta_{\alpha\beta}^{-1}(p),\label{glSTIwP}\\\
    & q^\mu\g_{\mu}(q,r,p) +\Cgh(q,r,p) = iD^{-1}(r^2) - iD^{-1}(p^2). 
    \label{ghSTIwP}
\end{align}
Taking the limit of~\2eqs{glSTIwP}{ghSTIwP} as $q\to 0$ on both sides, 
matching the zeroth order in $q$ yields the conditions
\begin{align}
    \widetilde{C}_{\alpha\beta}(0,r,-r)&=0;&
    \Cgh(0,r,-r)&=0,
    \label{zerothC}
\end{align}
whereas the terms linear in $q$ furnish a modified set of WTIs, namely
\begin{align}
	&\g_{\mu\alpha\beta}(0,r,-r) = -i\frac{\partial}{\partial r^\mu}\Delta^{-1}_{\alpha\beta}(r) - \left\lbrace\frac{\partial}{\partial q^\mu}\widetilde{C}_{\alpha\beta}(q,r,-r-q)\right\rbrace_{q=0}, \label{WI2glwithpole}\\  
    &\g_\mu(0,r,-r)  = -i\frac{\partial}{\partial r^\mu} D^{-1}(r^2)- \left\lbrace\frac{\partial}{\partial q^\mu}\Cgh(q,r,-r-q)\right\rbrace_{q=0}.
    \label{WI2ghwithpole}
\end{align}

The presence of the second term on the r.h.s. of \2eqs{WI2glwithpole}{WI2ghwithpole} has far-reaching consequences for the infrared behavior of $\Delta(q^2)$. Specifically, a repetition of the steps leading to \1eq{seag1} reveals that, whereas the first terms on the r.h.s. of these equations reproduces again \1eq{seag1} (and their contributions thus vanish), the second terms survive, giving 
\begin{align}
    \Delta^{-1}(0)&=\frac32g^2C_AF(0)\left\{\int_k k^2 \Delta^2(k^2)\left[1-\frac32g^2C_AY(k^2)\right]\Cgl'(k^2) -\frac13\int_k k^2 D^2(k^2)\Cgh'(k^2)\right\},
    \label{DSEmass}
\end{align}
where $C_A$ is the Casimir eigenvalue of the adjoint representation [$N$ for SU($N$)], $\Cgl$ is the form factor of $g_{\alpha\beta}$ in the tensorial decomposition of $\widetilde{C}_{\alpha\beta}$, and 
\begin{align}
    C_i^{\prime}(k^2)=\lim_{q\to0}\left\lbrace\frac{\partial \widetilde{C}_i(q,k,-k-q)}{\partial (k+q)^2}\right\rbrace,\quad i=\mathrm{gl},\ \mathrm{gh}.
\end{align}
As we see from \1eq{DSEmass}, a necessary condition for $\Delta^{-1}(0)$ to acquire a nonvanishing value is that at least one of the $\Cgl^{\prime}$ and $\Cgh^{\prime}$ does not vanish identically; in addition,  $\Cgl^{\prime}$ and $\Cgh^{\prime}$ must decrease sufficiently rapidly in the ultraviolet, in order for the integrals in  \1eq{DSEmass} to give a (positive) finite value. 

Let us conclude this section by linking the non-vanishing of $\Cgl^{\prime}$ to the generation of a running gluon mass of the type familiar from the quark case~\cite{Cloet:2013jya}. The infrared saturation of the gluon propagator suggests the physical parametrization  $\Delta^{-1}(q^2) =  q^2 J(q^2) + \hh (q^2)$ where $J(q^2)\sim\ln q^2$ at most, and $\hh (0)\neq 0$. Then the modified gluon STI~\noeq{glSTIwP} will make it natural to associate the $J$ terms with the $q^\mu\g_{\mu\alpha\beta}$ on the left-hand side (l.h.s.), and, correspondingly, 
\begin{align}
	\widetilde{C}_{\alpha\beta}(q,r,p) &= \hh(p^2) P_{\alpha\beta}(p) - \hh(r^2) P_{\alpha\beta}(r).
    \label{thectilde}
\end{align}
Focusing on the $g_{\alpha\beta}$ components of \1eq{thectilde},
we obtain~\cite{Aguilar:2011xe} 
\begin{align}
\Cgl(q,r,p)=\hh(r^2)-\hh(p^2)\quad \underset{q\to0}{\Longrightarrow}\quad 
\Cgl'(r^2)=
\frac{\diff m^2(r^2)}{\diff r^2}.
\label{theCgl}
\end{align}    
Then, upon integration, we obtain
\begin{align}
    \hh(q^2)=\Delta^{-1}(0)+\int_0^{q^2}\!\!\diff y\,\Cgl^\prime(y),
    \label{Cglvsmass}
\end{align}
thus establishing the announced link between $\Cgl^\prime$ and a dynamically generated gluon mass~\cite{Binosi:2012sj}.

\section{\label{sec:taylor} Taylor's theorem for the PT-BFM vertex $\widetilde{\Gamma}_\mu(q,r,p)$}

Taylor's theorem~\cite{Taylor:1971ff}, which is particular to the Landau gauge, establishes an exact constraint on the form factors comprising the conventional ghost-gluon vertex  (all momenta entering as usual)
\begin{align}
    i{\Gamma}_{c^nQ^a_\mu\bar c^m}(p,q,r)&=gf^{amn}{\Gamma}_\mu(q,r,p);&
    {\Gamma}^{(0)}_\mu(q,r,p)&=-r_\mu,
\end{align}
in the limit of vanishing ghost momentum ($p=0$). In this section, after briefly recalling how this theorem follows directly from the SDE satisfied by ${\Gamma}_\mu$, we derive the analogous relation for the BFM vertex
\begin{align}
    i{\Gamma}_{c^nB^a_\mu\bar c^m}(p,q,r)&=gf^{amn}\widetilde{\Gamma}_\mu(q,r,p);&
    \widetilde{\Gamma}^{(0)}_\mu(q,r,p)&=(p-r)_\mu,
\end{align}
using three different methods: \n{i} the Abelian STI~\noeq{AbWI3gh}, \n{ii} the BQI that connects ${\Gamma}_\mu$ with $\widetilde{\Gamma}_\mu$, and \n{iii} the SDE satisfied by $\widetilde{\Gamma}_\mu$. 

\subsection{Taylor's theorem for ${\Gamma}_\mu(q,r,p)$} 

\begin{figure}[!t]
    \centering
    \includegraphics[scale=0.65]{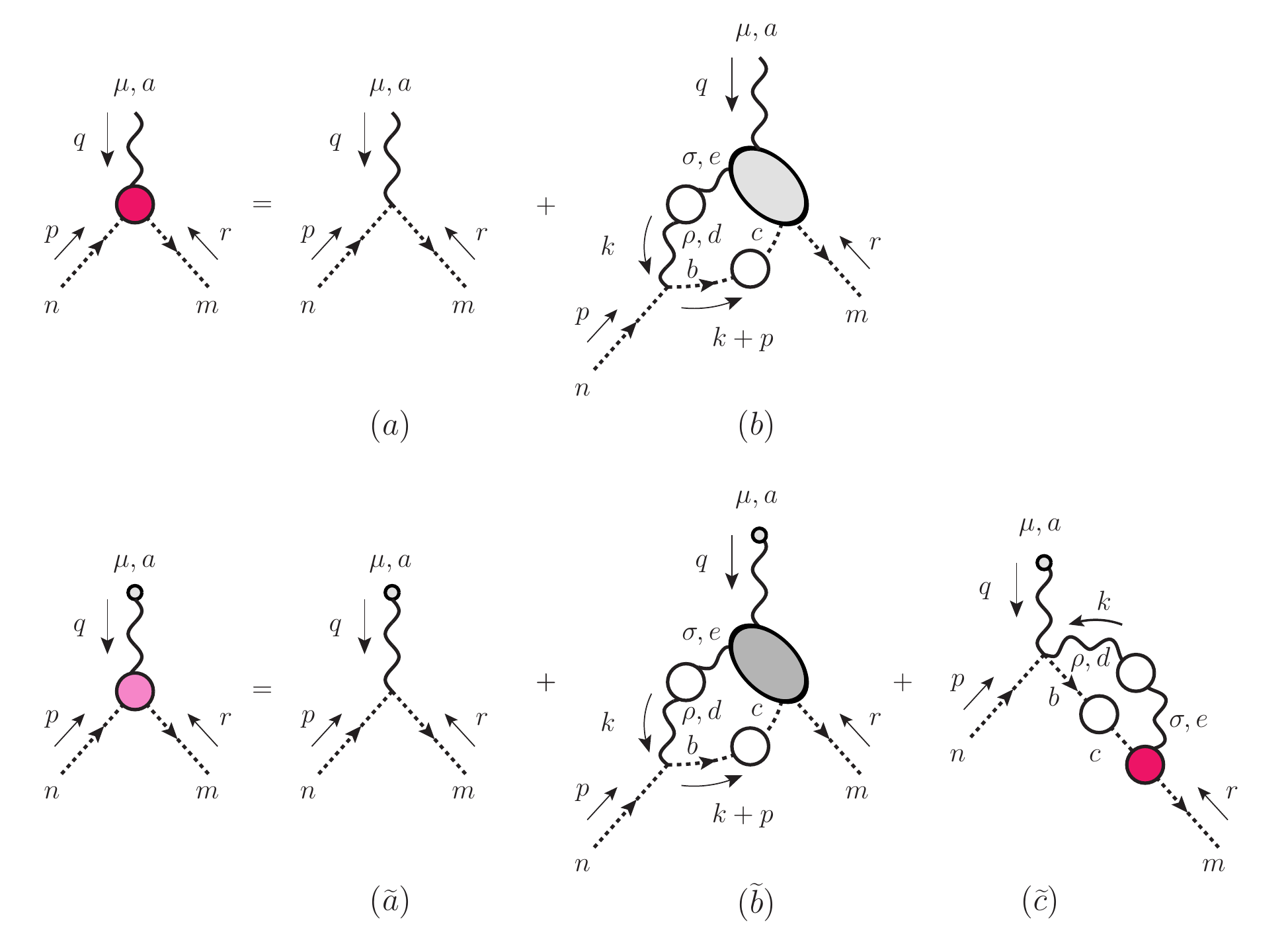}
    \caption{\label{fig:gluon-ghost-sde}The SDE satisfied by the gluon-ghost conventional (top) and BFM vertex (bottom). In this latter case an extra term $(\widetilde{c})$ appears, due to the presence of the additional BFM tree-level coupling $BQ c{\bar c}$.}
\end{figure}

The most compact version of Taylor's theorem may be obtained by using the gluon and ghost momenta ($q$ and $p$, respectively) for the tensorial decomposition of ${\Gamma}_\mu$, namely
\be 
{\Gamma}_\mu(q,r,p) = A(q,r,p) q_\mu+B(q,r,p) p_\mu.
\label{AB}
\ee
From the SDE of \fig{fig:gluon-ghost-sde}, we have that
\begin{align}
    gf^{amn}{\Gamma}_\mu(q,r,p) = -gf^{amn}r_\mu + gf^{dbn}\int_{k}(k+p)_\rho \Delta^{\rho\sigma}(k) D(k+p) {\cal Q}^{damb}_{\sigma\mu}(-k,q,r,k+p),
    \label{SDEnor}
\end{align}
where ${\cal Q}^{damb}_{\sigma\mu}$ represents the $QQc\bar c$ kernel appearing in diagram $(b)$ of that figure.
Evidently, in the Landau gauge, $(k+p)_\rho \Delta^{\rho\sigma}(k) = p_\rho \Delta^{\rho\sigma}(k)$, so that the entire contribution from the second term in~\1eq{SDEnor} vanishes when $p \to 0$. Thus, in the Taylor limit, \1eq{SDEnor} yields simply
\be 
{\Gamma}_\mu(q,-q,0) = q_\mu,
\label{T1}
\ee
while, from \1eq{AB}, in the same limit, we have that 
\be 
{\Gamma}_\mu(q,-q,0) = A(q,-q,0) q_\mu.
\label{T2}
\ee
Therefore, from \2eqs{T1}{T2} one obtains the known result 
\be 
A(q,-q,0) = 1. 
\label{Tcomp}
\ee

Notice that if instead one expresses ${\Gamma}_\mu(q,r,p)$ in terms of $q$ and $r$, namely
\be 
{\Gamma}_\mu(q,r,p) = A_1(q,r,p) q_\mu - B_1(q,r,p) r_\mu,
\label{ABprim}
\ee
we have that $A(q,r,p) = A_1(q,r,p)+B_1(q,r,p)$ and $B(q,r,p)=B_1(q,r,p)$, so that~\1eq{Tcomp} yields
\be
A_1(q,-q,0)+B_1(q,-q,0) =1,
\ee
which is the form of the theorem employed in previous works~\cite{Aguilar:2009nf,Aguilar:2013xqa}. 

\subsection{Taylor's theorem for $\widetilde{\Gamma}_\mu(q,r,p)$ from its STI}

Let us now turn to the vertex $\widetilde{\Gamma}_\mu(q,r,p)$, and consider its tensorial decomposition analogous to \1eq{AB}, 
\be 
{\widetilde\Gamma}_\mu(q,r,p) ={ \widetilde A}(q,r,p) q_\mu + {\widetilde B}(q,r,p) p_\mu.
\label{tildeAB}
\ee
Taking the limit $p \to 0$ we have 
\be 
{\widetilde\Gamma}_\mu(q,-q,0) = {\widetilde A}(q,-q,0) q_\mu,
\label{T2tilde}
\ee
and after contracting both sides by $q^{\mu}$ one gets
\be
q^\mu{\widetilde\Gamma}_\mu(q,-q,0) =  q^2 {\widetilde A}(q,-q,0). 
\label{L2}
\ee
On the other hand, from the STI we find
\be
q^\mu \widetilde{\Gamma}_\mu(q,r,p) = iD^{-1}(r^2) - iD^{-1}(p^2),
\label{L3}
\ee
which, as $p \to 0$, gives
\be
q^\mu{\widetilde\Gamma}_\mu(q,-q,0) =  q^2 F^{-1}(q^2).
\label{L4}
\ee
Thus, by combining \1eq{L2}  with \1eq{L4}, one obtains
\be
{\widetilde A}(q,-q,0) = F^{-1}(q^2),
\label{TTBFM}
\ee
which represents Taylor's theorem for the BFM ghost-gluon vertex.

\subsection{Derivation from the SDE}

We start by writing down the Landau gauge SDE for the ghost dressing function,
\be
 F^{-1}(q^2) = 1 + \Sigma (q^2),
\label{FSDE}
 \ee
where
\be
\Sigma (q^2) = i g^2  C_A \frac{q^{\mu}}{q^2} \int_{k}\!\Delta^{\mu\nu} (k) D(k+q) \Gamma_\nu(-k,-q,k+q).
\label{Sigma}
\ee

Next, let us consider the diagrammatic representation of the SDE satisfied by $\widetilde{\Gamma}_\mu(q,r,p)$, shown in~\fig{fig:gluon-ghost-sde}. 
The main subtlety in dealing with this SDE in the present context is the fact that its Landau gauge limit needs to be determined with particular care in the presence of diagrams containing the tree-level vertex $BQ^2$
\begin{align}
    &i\Gamma_{B^a_\mu Q^m_\alpha Q^n_\beta}(q,r,p)=gf^{amn}\widetilde{\Gamma}_{\mu\alpha\beta}(q,r,p),\nonumber \\
    &\widetilde{\Gamma}^{(0)}_{\mu\alpha\beta}(q,r,p)=g_{\alpha\beta}(r-p)_\mu+g_{\mu\beta}(p-q+\xiQ^{-1}r)_\alpha+g_{\mu\alpha}(q-r-\xiQ^{-1}p)_\beta.
\end{align}
As the above equation shows, this vertex differs from the corresponding  tree-level $Q^3$  vertex by a longitudinal term proportional to $1/\xiQ$, {\it i.e.}, 
\begin{align}
    \widetilde{\Gamma}^{(0)}_{\mu\alpha\beta}(q,r,p) &= {\Gamma}^{(0)}_{\mu\alpha\beta}(q,r,p) -
    \xiQ^{-1}\Gamma^\s{\mathrm{P}}_{\mu\alpha\beta}(q,r,p);&
    \Gamma^\s{\mathrm{P}}_{\mu\alpha\beta}(q,r,p)&=
    p_{\beta} g_{\mu\alpha} - r_{\alpha} g_{\mu\beta}. 
\end{align}

This implies in turn that, as has been explained in~\cite{Aguilar:2008xm}, the limit $\xiQ \to 0$ must be achieved by letting each of the longitudinal momenta act on the adjacent gluon propagator (written for a general $\xi$), yielding, \eg $p_{\beta}\Delta^{\beta\rho} (p) =  -i\xiQ p^{\rho}/p^2$; in this way the would-be divergent $\frac{1}{\xiQ}$ is cancelled out, and one may set directly $\xiQ = 0$ in the remaining expression. 

\begin{figure}[!t]
    \centering
    \includegraphics[scale=0.6]{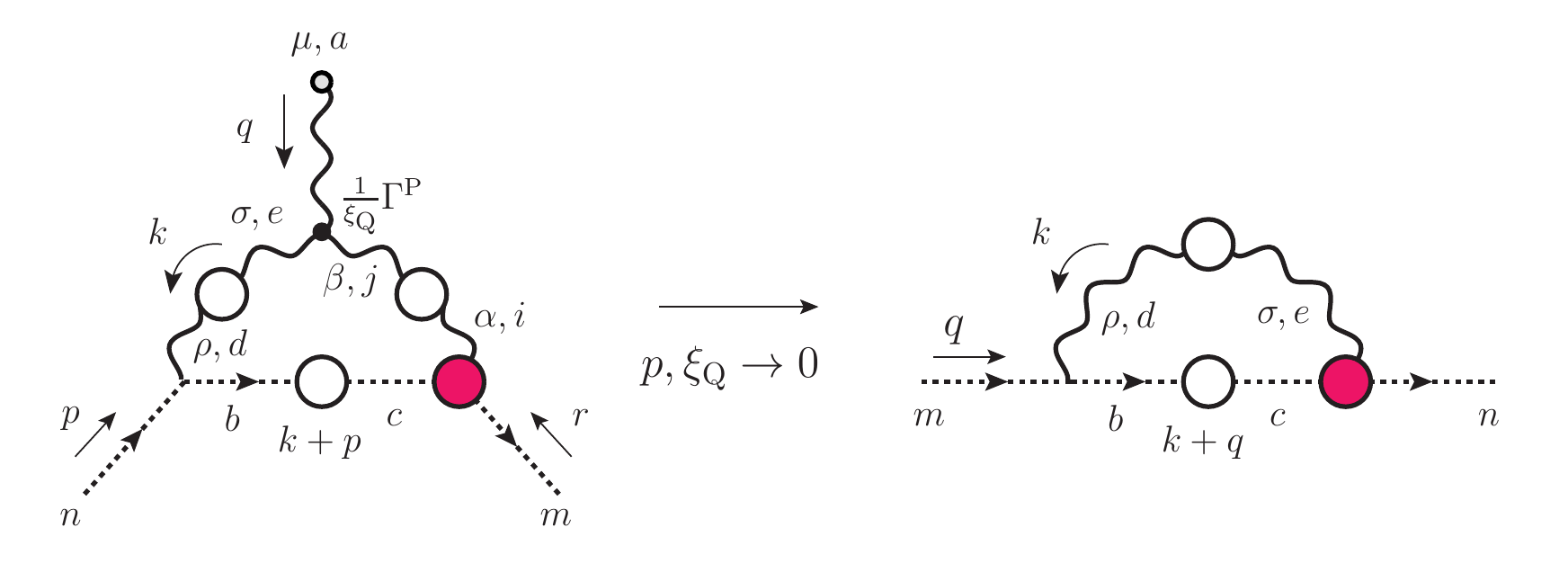}
    \caption{\label{fig:gammaP}The unique contribution to the $(\widetilde b)$ diagram of~\fig{fig:gluon-ghost-sde} which is non-vanishing in the $p\to0$ limit.}
\end{figure}

These observations  are particularly relevant when evaluating  diagram $(\widetilde b)$ of~\fig{fig:gluon-ghost-sde}, because, unlike its counterpart $(b)$, it does {\it not} vanish in the limit $p \to 0$. The easiest way to appreciate this fact it to remember that the vanishing of graph $(b)$ relies on the fact that the term $(k+p)_{\rho}$ originating from the tree-level ghost-gluon vertex is contracted  with an adjacent  $\Delta^{\rho\sigma} (k)$ in the Landau gauge, see~\1eq{SDEnor}. However, if the 
$\Delta^{\rho\sigma}$ enters, in its other end, into a  
tree-level vertex $\widetilde{\Gamma}^{(0)}$, the longitudinal momentum $(k+p)_\sigma$ present in $\Gamma^\s{\mathrm{P}}$  will act on it; thus, the original $(k+p)_{\rho}$ will be contracted with $(k+p)^{\rho}/(k+p)^2$ instead, 
and will therefore survive when the limit $p\to0$ 
is taken.

It turns out that there is only one possible structure of this type contained in $(\widetilde b)$, which is shown diagrammatically in~\fig{fig:gammaP}; then, it is relatively straightforward to establish that, in the $p\to0$ limit, we have that
\begin{align}
    (\widetilde b)_\mu = \frac{1}{2} q_\mu \Sigma (q^2),
\end{align}
with the 1/2 factor originating from the use of the 
identity $f^{ads} f^{msb} f^{nbd} = \frac{1}{2} C_A f^{amn}$.

Finally, one needs to consider the additional diagram $(\widetilde{c})$ which appears due to the presence of the PT-BFM special vertex $BQc {\bar c}$ 
\begin{align}
    \Gamma^{(0)}_{c^nB^a_\mu Q^b_\nu\bar c^m}(p,q,t,r)=-ig^2g_{\mu\nu}f^{mae}f^{ebn}.
\end{align}
In the $p\to0$ limit then one obtains for this diagram
\be
(\widetilde c)_\mu = \frac{1}{2} q_\mu \Sigma (q^2),
\ee
which, when added to the previous result, gives for the $\widetilde{\Gamma}_\mu$ SDE in the $p\to0$ limit  
\be
\widetilde{\Gamma}_\mu(q,-q,0) =  q_{\mu} [1 + \Sigma (q^2)]=q_{\mu}F^{-1}(q^2),
\label{BQIr0}
\ee
where~\1eq{FSDE} has been used.

\subsection{Derivation from the BQI}

\begin{figure}[!t]
    \centering
    \includegraphics[scale=0.65]{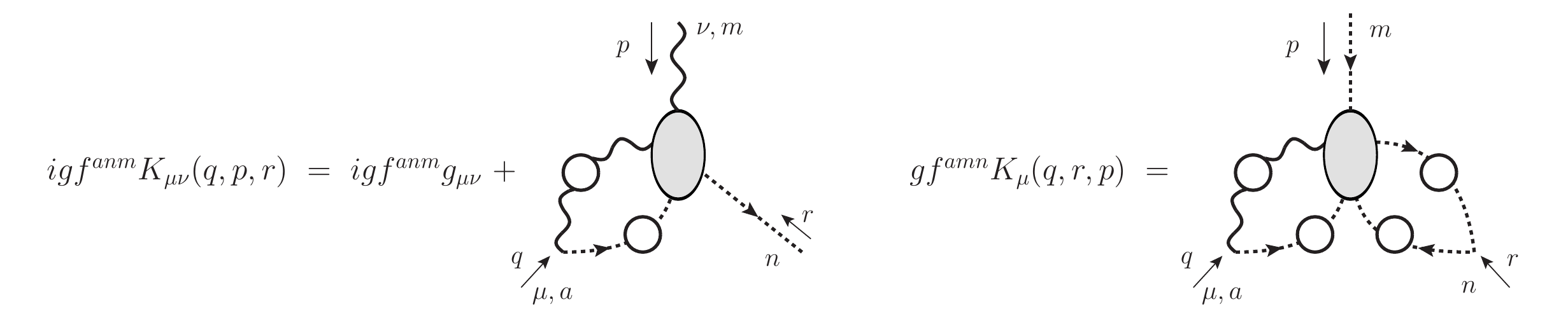}
    \caption{\label{fig:BQIaux}The auxiliary functions appearing in the ghost-gluon vertex BQI.}
\end{figure}

Finally, let us consider the BQI that relates the conventional and background ghost-gluon vertices, which reads~\cite{Binosi:2008qk}
\begin{align}
    \widetilde{\Gamma}_\mu(q,r,p) &=\left\{[1+G(q^2)]g_{\mu}^{\nu} + \frac{q_{\mu} q^{\nu}}{q^2} L(q^2) \right\}{\Gamma}_\nu(q,r,p) + F^{-1}(p^2) p^{\nu} K_{\mu\nu}(q,p,r)\nonumber\\
    &- r^2  F^{-1}(r^2) K_{\mu}(q,p,r),
    \label{BQI}
\end{align}
where $K_{\mu\nu}$ and $K_{\mu}$ are the auxiliary Green's functions shown in~\fig{fig:BQIaux}, which involve composite operators appearing as a consequence of the anti-BRST symmetry present when quantizing the theory within the BFM framework~\cite{Binosi:2013cea}.

When taking the $p\to0$ limit, on the one hand the second term on the right-hand side (r.h.s.) of the BQI~\noeq{BQI} vanishes directly due to the presence of $p^{\nu}$; on the other hand, the last term vanishes in the Landau gauge, because the relation $(k+p)_\rho \Delta^{\rho\sigma}(k) = p_\rho \Delta^{\rho\sigma}(k)$ will be triggered once again. Thus, in this limit, the BQI reduces to 
\begin{align}
    \widetilde{\Gamma}_\mu(q,-q,0) =  \left\{[1+G(q^2)]g_{\mu}^{\nu} + \frac{q_{\mu} q^{\nu}}{q^2} L(q^2) \right\}{\Gamma}_\nu(q,-q,0).
\end{align}
Now, Taylor's theorem for the conventional vertex implies ${\Gamma}_\nu(q,-q,0) = q_\nu$, so that one arrives at 
\begin{align}
    \widetilde{\Gamma}_\mu(q,-q,0) = \left[1+G(q^2)+L(q^2)\right]q_\mu \,.
\end{align}
At this point, use of the Landau gauge relation~\cite{Grassi:2004yq,Aguilar:2009nf}
\begin{align}
    F^{-1}(q^2)=1+G(q^2)+L(q^2),
    \label{funrel}
\end{align}
together with \1eq{T2tilde}, leads immediately to the result of \1eq{TTBFM}.

\section{\label{sec:polepart}A closer look at the pole part of the ghost vertex}

It is well understood that, in order for the gluon mass generation to go through
in the  way described in~\cite{Aguilar:2011xe,Ibanez:2012zk,Aguilar:2016vin}, the  STIs satisfied by
the  fundamental vertices  must  be realized  in part  by  means of  a
longitudinally coupled pole term.  This fact, in turn, imposes general
restrictions on the structure of the form factors of these vertices; in this
section we will study this issue for the case of the $Bc\bar c$ vertex  $\widetilde{\Gamma}_\mu$, which, due to its reduced tensorial content, is particularly instructive. 
In the first subsection we examine in some  detail the structure of the pole part
of $\widetilde{\Gamma}_\mu$, its relation with the other form factors,
together with the restrictions imposed by Taylor's theorem.
Then, in the second subsection, we introduce a concrete {\it Ansatz} for
the pole part, which, in conjunction with the 
solution obtained from the BSE system in Sec.~\ref{numan}, 
allows for the  sequential determination of all relevant pieces of  $\widetilde{\Gamma}_\mu$.

\subsection{General considerations and alternative formulation}

We start by considering the general form of the vertex $\widetilde{\Gamma}_\mu(q,r,p)$,
given by 
\begin{align}
	{\widetilde\Gamma}_\mu(q,r,p) &={\widetilde A}^{\rm np}(q,r,p) q_\mu + {\widetilde B}^{\rm np}(q,r,p) p_\mu+\frac{q_\mu}{q^2} \Cgh(q,r,p),\label{wGfirst}
\end{align}
where both ${\widetilde A}^{\rm np}$ and ${\widetilde B^{\rm np}}$ are finite functions for all possible momenta $q$, $r$, and $p$. If we now take the limit $p\to 0$ on the r.h.s. of \1eq{wGfirst} and use Taylor's theorem, we conclude that ${\widetilde A}^{\rm np}(q,-q,0)$ and $\Cgh(q,-q,0)$ must satisfy the constraint 
\begin{align}
	\Cgh(q,-q,0) + q^2 {\widetilde A}^{\rm np}(q,-q,0) = q^2 F^{-1}(q^2).
	\label{Gen1}
\end{align}
Note that, since $F^{-1}(q^2)$ and ${\widetilde A}^{\rm np}(q,-q,0)$ are finite at the origin, \1eq{Gen1} implies that $\Cgh(0,0,0) = 0$ [this last result may be obtained also from by setting $r=0$ directly in the condition~\noeq{zerothC}]. 

Let us now introduce 
\begin{align}
	{\cal R}(q,r,p) := i\frac{D^{-1}(r^2) - D^{-1}(
	p^2)}{r^2-p^2} = \frac{r^2 F^{-1}(r^2) - p^2 F^{-1}(p^2)}{r^2-p^2},
	\label{Rdef}
\end{align}
and, without loss of generality, set
\begin{align}
	{\widetilde A}^{\rm np}(q,r,p) &= {\cal R}(q,r,p) + f_{\s A}(q,r,p),\nonumber\\
	{\widetilde B^{\rm np}}(q,r,p) &=  2\, {\cal R}(q,r,p) + f_{\s B}(q,r,p),
	\label{ABf}
\end{align}
where $f_{\s A}$ and $f_{\s B}$ are arbitrary, purely non-perturbative functions, assumed to be  well-behaved in the entire range of their arguments, and in particular in the important limits $q\to 0$ and $p\to 0$. Note that the tree-level values for ${\widetilde A}^{\rm np}$ and ${\widetilde B}^{\rm np}$ are correctly recovered, since ${\cal R}^{(0)}=1$. 

Evidently, \1eq{Rdef} implies ${\cal R}(q,-q,0)= F^{-1}(q^2)$; therefore
\begin{align}
	{\widetilde A}^{\rm np}(q,-q,0)= F^{-1}(q^2) + f_{\s A}(q,-q,0),
\end{align} 
and from~\1eq{Gen1} we must have that
\begin{align}
\Cgh(q,-q,0) = -q^2 f_{\s A}(q,-q,0).
\label{CfA}
\end{align}

Let us next contract $\widetilde{\Gamma}_\mu(q,r,p)$ by $q^{\mu}$; clearly, the terms proportional to ${\cal R}(q,p,r)$ saturate the STI, and thus we must have 
\begin{align}
q^2 f_{\s A}(q,r,p) + (p\cd q) f_{\s B}(q,r,p) + \Cgh(q,r,p) = 0.  
\label{constr}
\end{align}
Note that, in the limit $p\to 0$, \1eq{constr} simply reproduces \1eq{CfA}; however, if we take instead the limit $q\to 0$, the matching of the linear terms in $q$ yields the additional relation 
\begin{align}
f_{\s B}(0,r,-r) = 2 \Cgh^{\prime}(r^2).
\label{fBC}
\end{align}
This relation is particularly interesting because it connects {\it explicitly} the term $\Cgh^{\prime}(r^2)$ that accompanies the massless pole (and enters  eventually in the ``mass'' equation (\ref{DSEmass})) with the function $f_{\s B}$, which quantifies the necessary deviation of ${\widetilde B^{\rm np}}(q,r,p)$ from the expression that would saturate the STI identically. At this point one may verify immediately that, as first stated in~\cite{Aguilar:2016vin} (see Eq.~(7.4) there\footnote{Notice that the form factor ${\cal A}^{\rm{np}}_2$ defined in~\cite{Aguilar:2016vin} carries in the $q\to0$ limit a minus sign with respect to the ${\widetilde B}^{\rm np}$ defined here, see~Eqs.~(3.17) and~(3.18) in~\cite{Aguilar:2016vin}.}), 
\begin{align}
{\widetilde B^{\rm np}}(0,r,-r) = 2 \left[ i\frac{\partial {D}^{-1} (r^2)}{\partial r^2} + \Cgh^{\prime}(r^2)\right].
\end{align}

It is evident from the above considerations,
and particularly from \1eq{fBC}, that the terms
of $\widetilde{\Gamma}_\mu(q,r,p)$ that involve $f_{\s A}$, $f_{\s B}$, and
$\Cgh$ must organize themselves into a transverse structure. 
To see this explicitly, use \1eq{constr} to eliminate 
any of the $\Cgh$, $f_{\s A}$ and $f_{\s B}$ in favor of the other two,  
and substitute into \1eq{wGfirst}, to obtain 
\begin{align}
   {\widetilde\Gamma}_\mu(q,r,p) = (2p+q)_\mu {\cal R}(q,r,p)
   + f_{\s B}(q,r,p)\, p^{\sigma} P_{\sigma\mu} (q).
\label{Ans1}
\end{align}
Clearly, the expression on the r.h.s. of \1eq{Ans1} yields directly the correct Taylor limit. Note also that $q^2 p^{\sigma} P_{\sigma\mu} (q) = (p\cd q)\, r_\mu - (r\cd q)\, p_\mu$, the latter being the transverse vector introduced by Ball and Chiu\footnote{The vertex studied in~\cite{Ball:1980ay} is not 
$\widetilde{\Gamma}_\mu$, but rather 
the photon-scalar vertex of scalar QED. However, apart from the overall color factor, 
there is a direct one-to-one correspondence between the two vertices,  
mainly due to the fact that they both satisfy a similar Abelian STI, 
namely that of \1eq{AbWI3gh}, 
with the simple replacement $D(q^2)\to {\cal D}(q^2)$, where  
${\cal D}(q^2)$ is the propagator of the charged scalar particle.}~\cite{Ball:1980ay}.

According to \1eq{Ans1}, all memory of the longitudinally coupled pole has been transferred to the transverse part of the vertex. Of course, this simple reorganization of terms leading to  \1eq{Ans1} could not possibly induce any modifications to the contribution of the ghost loops to $\Delta^{-1}(0)$. To see that this is indeed so, note that the first term of \1eq{Ans1}, in the limit $q \to 0$, triggers the ``seagull identity'' and cancels exactly against the seagull diagram, while the second term gives a contribution that is {\it manifestly transverse} ($p \to k$), 
\begin{align}
P_{\mu\nu}(q) \widetilde\Pi (q^2) =  g^2 C_A P_{\sigma\nu} (q)
\int_{k}k_\mu k^{\sigma} D(k) D(k-q) f_{\s B}(q,k-q,-k).
\end{align}
Then, as $q\to 0$, we obtain 
\begin{align}
\widetilde\Pi (0) = \frac{g^2 C_A}{d} \int_{k} k^{2} D^{2}(k) f_{\s B}(0,k,-k),
\label{mgh}
\end{align}
which, after taking into account \1eq{fBC}, coincides with Eq.~(6.11) of~\cite{Aguilar:2016vin} (see also Eq.~(7.3) of the same paper).

Let us point out that the ${\widetilde\Gamma}_\mu$ of \1eq{Ans1} could have been supplemented from the beginning by a transverse piece, whose form factor, unlike that of \1eq{Ans1}, would vanish as $q\to 0$; this is indeed the construction of~\cite{Ball:1980ay}, where a term $a(q,r,p) \left[ (r\cd q)\, p_\mu - (p\cd q)\, r_\mu)\right]$ is included, with $a(q,r,p)$ finite. In the present context, the effect of including this additional term would be to modify $f_{\s B} \to f_{\s B}+ q^2 a$; this extra term is clearly irrelevant as far as the gluon mass generation is concerned; for instance, it would have a vanishing contribution to the r.h.s. of \1eq{mgh}. Therefore, $a(q,r,p)$ will be neglected in what follows.

\subsection{A special case}

Let us now consider a special realization of the general scenario presented above, which admits a complete solution. Specifically, we set
\begin{align}
f_{\s A}(q,r,p) = f (q,r,p) = \frac{1}{2} f_{\s B}(q,r,p),
\label{z1}
\end{align}
which, using~\1eq{constr}, implies
\begin{align}
f(q,r,p) = -\frac{\Cgh(q,r,p)}{r^2-p^2}.
\label{z2}
\end{align}
Next, for $\Cgh$ we employ, similarly to what we were lead to in~\1eq{thectilde} for the gluon case, the simple {\it Ansatz}
\begin{align}
\Cgh(q,r,p) = r^2 h(r^2) - p^2 h(p^2),
\label{z3}
\end{align}
which clearly satisfies the condition $\Cgh(0,r,-r) =0$, as required on general grounds. In addition, the quantity $\Cgh^{\prime}(r^2)$ is now given by 
\begin{align}
\Cgh^{\prime}(r^2) = [r^2 h(r^2)]^{\prime},
\label{z4}
\end{align}
while, in the Taylor limit, 
\begin{align}
\Cgh(q,-q,0) = q^2 h(q^{2}) = -q^2 f (q,-q,0),
\label{z5}
\end{align}
exactly as required from \1eq{CfA}.

The above {\it Ansatz} allows for a complete solution of the part of the ghost sector that affects the dynamics of the gluon mass generation, because, once $\Cgh^{\prime}(r^2)$ has been determined from the corresponding BSE system, all other quantities may be deduced from~\1eq{ABf} together with \1eq{z1} through~\noeq{z4}. 

In particular,
\begin{align}
r^2 h(r^2)  = c + \int_0^{r^{2}}\!\diff y \, \Cgh^{\prime}(y) 
\label{z6}
\end{align}
where $c$ is the integration constant. Evidently, $c$ drops out when forming $\Cgh(q,r,p)$ using \1eq{z4}, 
\begin{align}
\Cgh(q,r,p) = \int_0^{r^{2}}\! \diff y \, \Cgh^{\prime}(y) - \int_0^{p^{2}} \diff y \, \Cgh^{\prime}(y);
\label{z7}
\end{align}
on the other hand, in the Taylor limit ($p \to 0$) \1eq{z7} yields
\begin{align}
\Cgh(q,-q,0) = \int_0^{q^{2}} dy \, \Cgh^{\prime}(y),
\label{z8}
\end{align}
which may be reconciled with \1eq{z5} and \1eq{z6} only for the value  $c=0$.

At this point it is natural to introduce the combination
\begin{align}
F_{{\rm eff}}^{-1}(q^2) &:= F^{-1}(q^2) + h(q^2) 
\nonumber\\
 &= F^{-1}(q^2)[1+ \underbrace{h(q^2) F(q^2)}_{\delta(q^2)}],
\end{align}
where the function
\begin{align}
\delta(q^2) = D(q^2) \int_0^{q^2} dy \, \Cgh^{\prime}(y) 
\label{thedelta}
\end{align}
quantifies the relative deviation of the  vertex form factors from their ``canonical'' form,    
due to the presence of the pole term. Specifically, one obtains
\begin{align}
{\widetilde A}^{\rm np}(q,r,p) = {\cal R}_{{\rm eff}}(q,r,p) =\frac{1}{2} {\widetilde B^{\rm np}}(q,r,p)\,,
\end{align}
where  ${\cal R}_{{\rm eff}}$ is obtained from the ${\cal R}$ in \1eq{Rdef}
by carrying out the substitution $F^{-1}(q^2) \to F_{{\rm eff}}^{-1}(q^2)$.

%%%%%%%%%%%%%%%%%%%%%%%%%%%%%%%%%%%%%%%%%%%%%%%%%%%%%%%%%%%%%%%%%%%%%%%%%%%

\section{\label{furcon}Coupled Dynamics of massless pole formation}

The actual behavior of  $\Cgl^{\prime}$ and $\Cgh^{\prime}$ is determined by a homogeneous system of linear integral equations, which may be derived from the SDEs satisfied by the corresponding $BQ^2$ and $Bc\bar c$ vertices as $q \to 0$~\cite{Aguilar:2011xe,Aguilar:2016ock}. As in this limit the zeroth order terms vanish by  virtue of~\1eq{zerothC}, the derivative terms become the leading contributions, and the resulting homogeneous equations assume the form of two coupled BSEs, given by
\begin{align}
    f^{amn}\lim_{q\to 0} \widetilde{C}_{\alpha\beta}(q,r,p)
    = f^{abc}\lim_{q\to 0} \Bigg\{ & \int_k \widetilde{C}_{\gamma\delta}(q,k,-k-q)\Delta^{\gamma\rho}(k)\Delta^{\delta\sigma}(k+q){\cal K}_{1\rho\alpha\beta\sigma}^{bmnc}(-k,r,p,k+q)\nonumber \\
    +&\int_k\Cgh(q,k,-k-q)D(k)D(k+q){\cal K}_{2\alpha\beta}^{bmnc}(-k,r,p,k+q)
    \Bigg\},\nonumber \\
    f^{amn}\lim_{q\to 0} \Cgh(q,r,p)
    = f^{abc}\lim_{q\to 0} \Bigg\{ & \int_k \widetilde{C}_{\gamma\delta}(q,k,-k-q)\Delta^{\gamma\rho}(k)\Delta^{\delta\sigma}(k+q){\cal K}_{3\rho\sigma}^{bmnc}(-k,r,p,k+q)\nonumber \\
    +&\int_k\Cgh(q,k,-k-q)D(k)D(k+q){\cal K}_{4}^{bmnc}(-k,r,p,k+q)
    \Bigg\}.\nonumber \\
\end{align}

\begin{figure}[!t]
    \centering
    \includegraphics[scale=0.5]{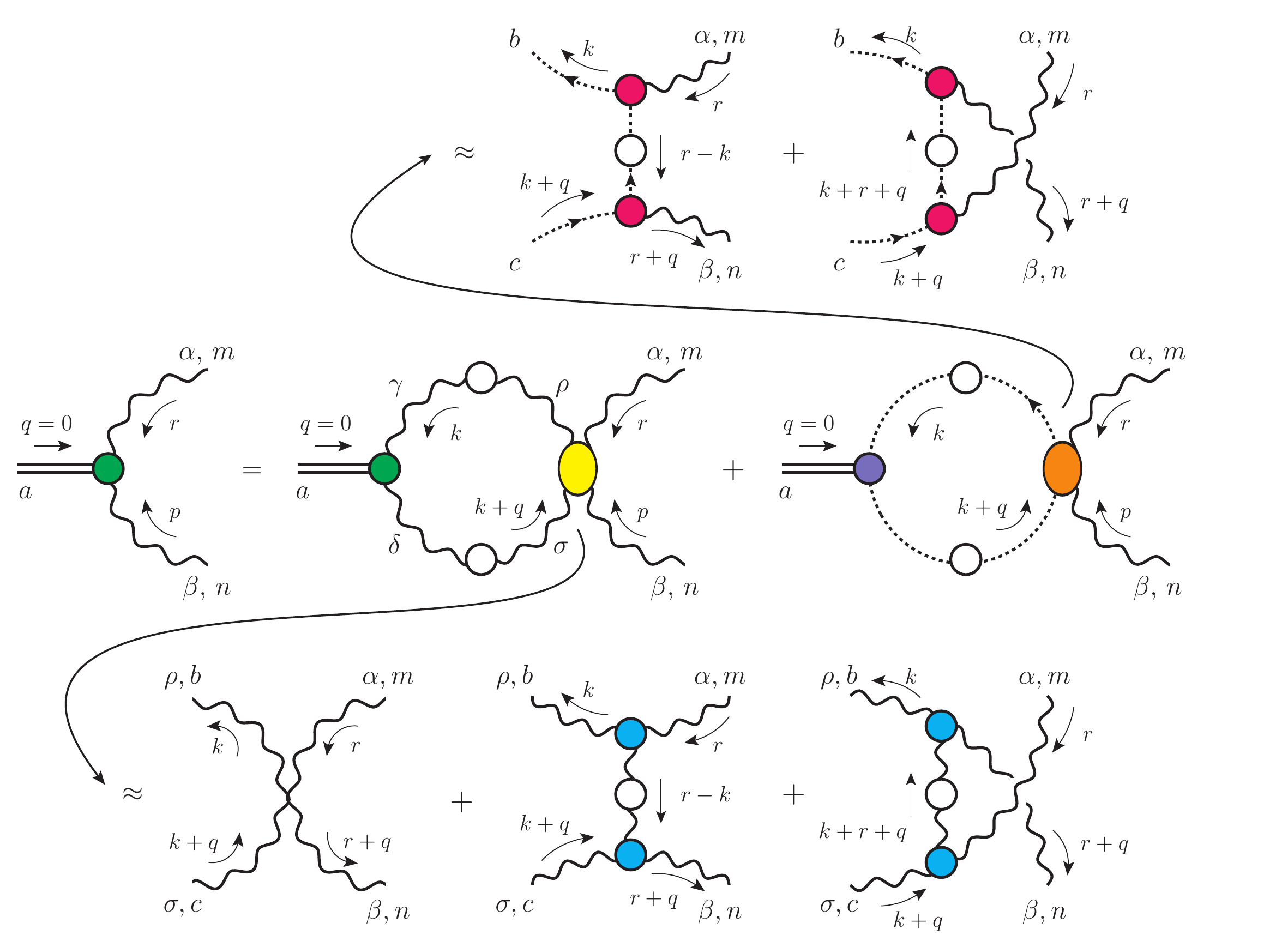}
    \caption{\label{fig:gl-bse}The BSE satisfied by the gluon bound-state wave function $\widetilde{C}_{\alpha\beta}$ (center) in the presence of both gluon and ghost massless poles. The simplified four-gluon and gluon-ghost kernels used are also shown.}
\end{figure}

\begin{figure}[!t]
    \centering
    \includegraphics[scale=0.5]{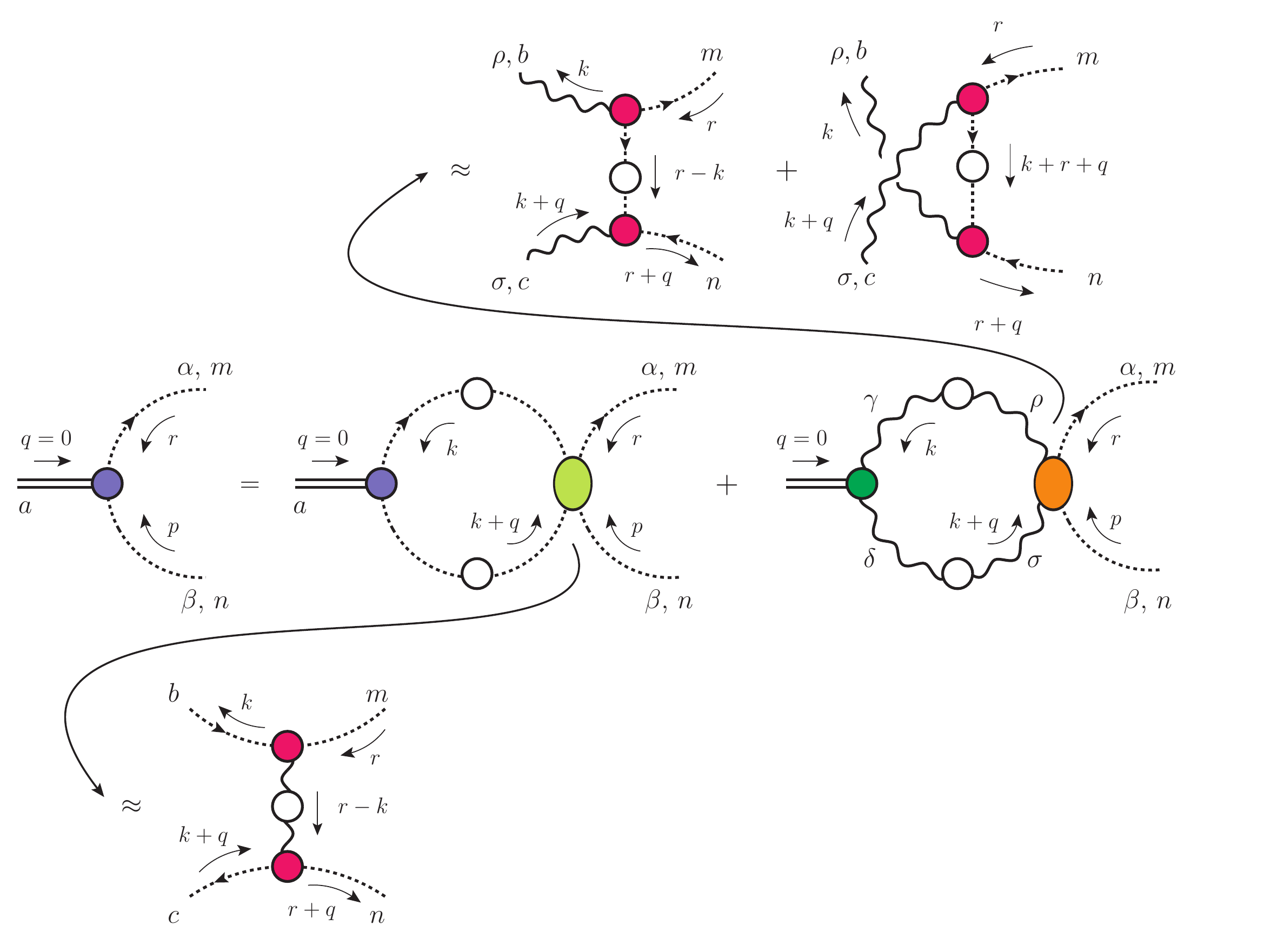}
    \caption{\label{fig:gh-bse}The BSE satisfied by the ghost bound-state wave function $\Cgh$ (center) in the presence of both gluon and ghost massless poles. The simplified four-ghost and gluon-ghost kernels used are also shown.}
\end{figure}

To proceed further, we will approximate the four-point BS kernels ${\mathcal K}_i$ by their lowest-order set of diagrams shown in \fig{fig:gl-bse} and \fig{fig:gh-bse}, in which the various diagrams contain fully dressed propagators and vertices (notice that all gluon propagators are ``quantum" ones, and all vertices
of the ``\,$\Gamma$ type"). In particular for the three-gluon and ghost-gluon vertices we will consider the simple {\it Ans\"atze}
\begin{align}
    \Gamma_{\mu\alpha\beta}(q,r,p)=
	\fgl (r)\Gamma^{(0)}_{\mu\alpha\beta}(q,r,p),\nonumber\\
	\Gamma_{\mu}(q,r,p)=
	\fgh (r)\Gamma^{(0)}_{\mu}(q,r,p),
\end{align}
where $\Gamma^{(0)}$ represents the standard tree-level expression of the corresponding vertex, and the form factors $\fgl$ and $\fgh$ are considered to be functions of a single kinematic variable. We then arrive at the following final equations
\begin{align}
    \Cgl^\prime(q^2)&=\frac{8\pi}3\alpha_sC_A\!\left[\kint\Cgl^\prime(k^2)\Delta^2(k)\Delta(k+q){\cal N}_1(k,q)+\frac14\kint\Cgh^\prime(k^2)D^2(k)D(k+q){\cal N}_2(k,q)\!
    \right]\!,\nonumber \\
    \Cgh^\prime(q^2)&=2\pi\alpha_sC_A\!\left[\kint\Cgl^\prime(k^2)\Delta^2(k)D(k+q){\cal N}_3(k,q)+\frac12\kint\Cgh^\prime(k^2)D^2(k)\Delta(k+q){\cal N}_4(k,q)
    \!\right]\!,
    \label{TheSys}
\end{align}
where
\begin{align}
{\cal N}_1(k,q)&=\frac{(q\!\cdot\!k)[q^2k^2-(q\!\cdot\! k)^2]}{q^4k^2(k+q)^2}\fgl^2(k+q)\left[8q^2k^2 + 6(q\!\cdot\!k)(q^2+k^2)+3(q^4+k^4)+(q\!\cdot\! k)^2\right],\nonumber \\
{\cal N}_2(k,q)&=\frac{(q\!\cdot\!k)[q^2k^2-(q\!\cdot\! k)^2]}{q^4}\fgh^2(k+q),\nonumber \\
{\cal N}_3(k,q)&=\frac{(q\!\cdot\!k)[q^2k^2-(q\!\cdot\! k)^2]}{q^2k^2}\fgh^2(k+q),\nonumber \\
{\cal N}_4(k,q)&=\frac{(q\!\cdot\!k)[q^2k^2-(q\!\cdot\! k)^2]}{q^2(k+q)^2}\fgh^2(k+q).
\end{align}
Notice, in particular, that in the $q\to0$ limit, $\Cgl(0)$ saturates to a constant~\cite{Binosi:2017rwj}, whereas the structure of the ${\cal N}_3$ and ${\cal N}_4$ kernels implies that $\Cgh(0)=0$.  

\begin{figure*}[!t]
    \centering
    \mbox{}\hspace{-.8cm}
    \includegraphics[scale=0.6]{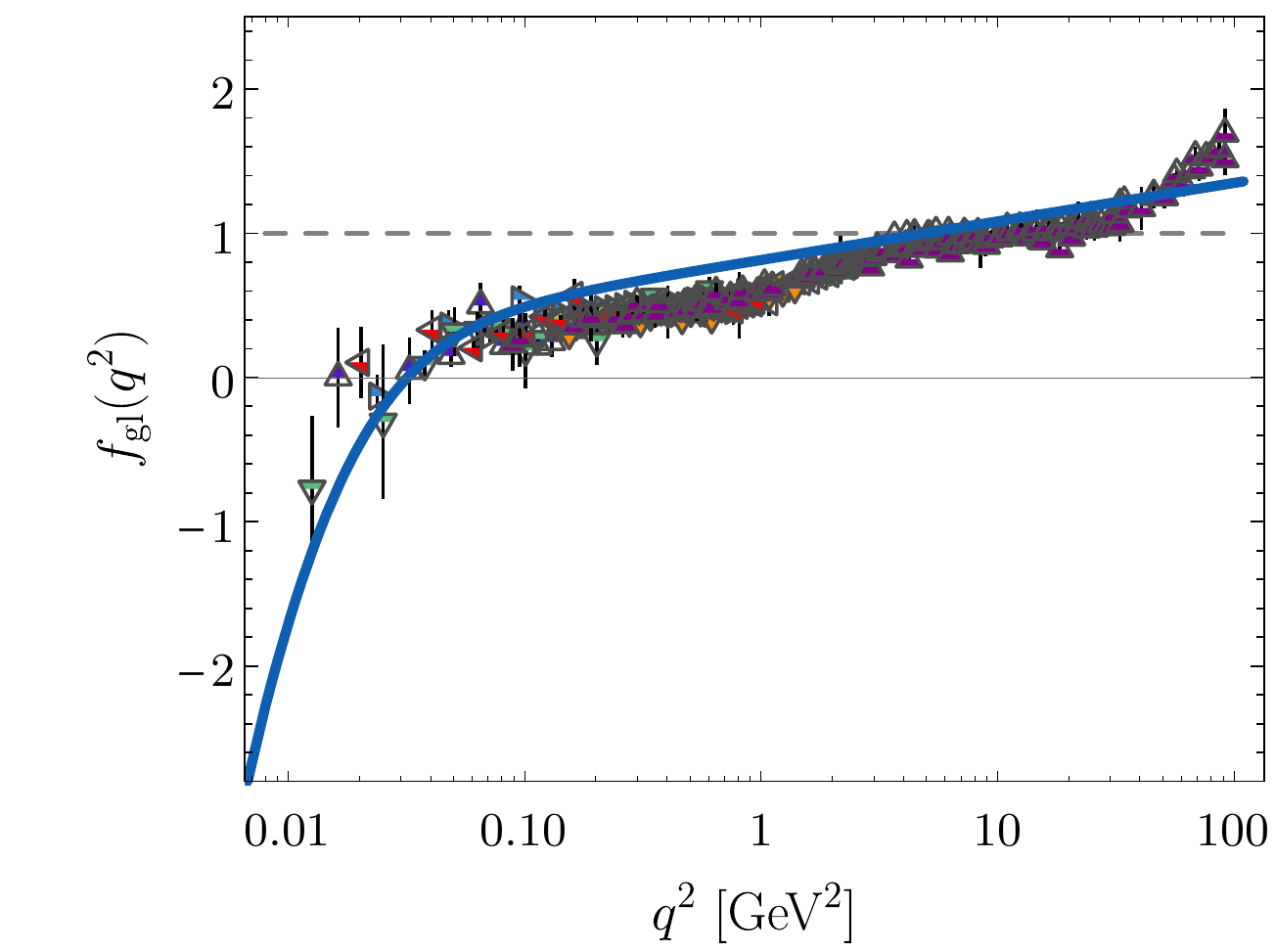}\hspace{.3cm}
    \includegraphics[scale=0.6]{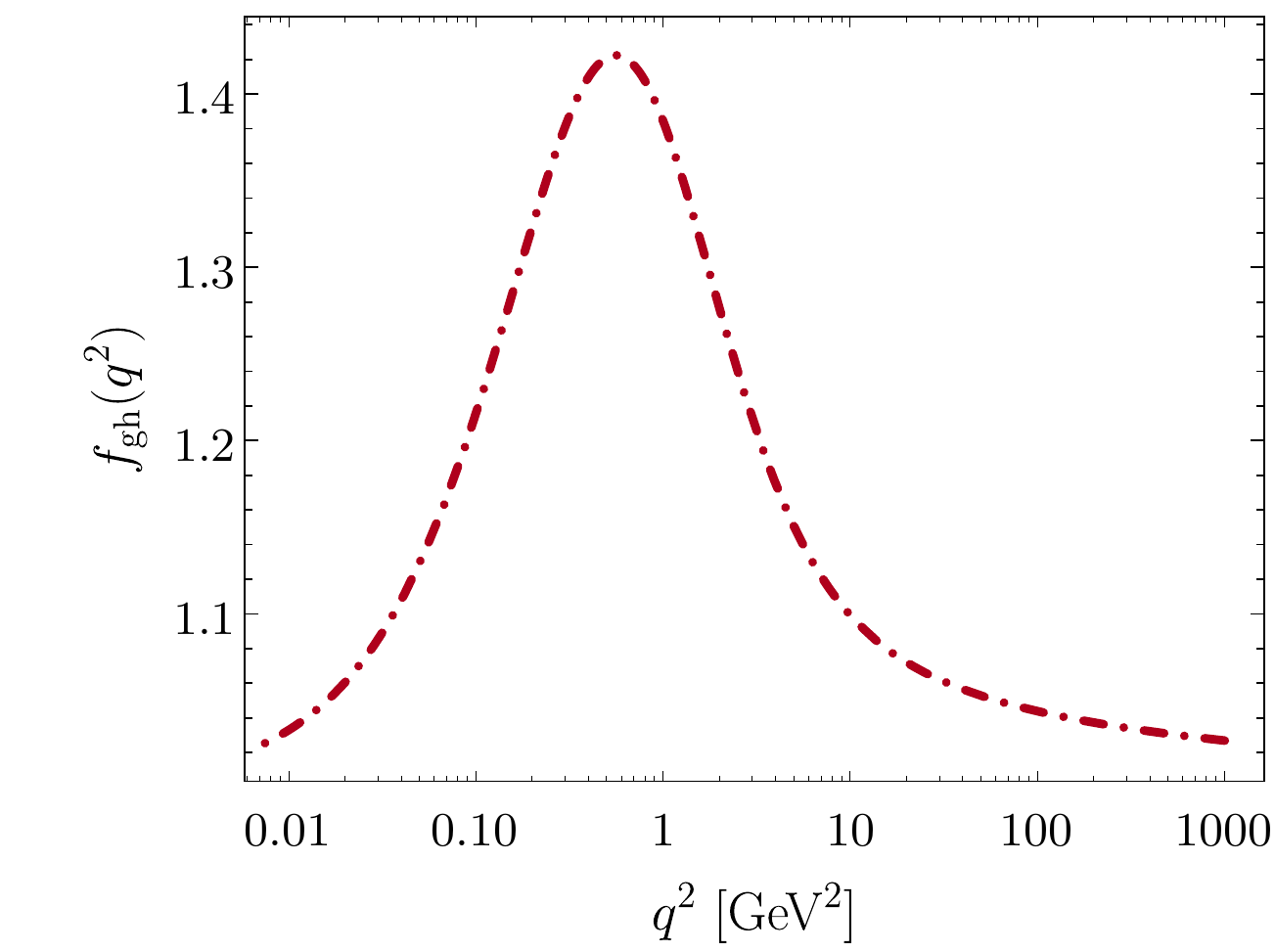}
    \caption{\label{fig:f}(Left panel) SU(3) lattice data (evaluated with various $\beta$, volumes and lattice actions) for the form factor $\fgl$ in the symmetric configuration~\cite{Athenodorou:2016oyh,Boucaud:2017obn}; the continuous line corresponds to the optimal data description obtained in~\cite{Binosi:2017rwj} when solving the BSE in the absence of ghosts. (Right panel) The ghost-gluon vertex form factor $\fgh$ in the symmetric configuration obtained from solving its SDE.}
\end{figure*}

\section{\label{numan}Numerical Analysis}

Before proceeding to solve the BSE system~\noeq{TheSys}, some of the functions that appears in it ought to be specified.

To begin with, for the gluon propagator $\Delta$ and ghost dressing function $F$ we will 
employ the available SU(3) lattice data~\cite{Bogolubsky:2009dc}.
As for the vertex form factors $\fgl$ and $\fgh$,
we use the curves shown in~\fig{fig:f}.
More specifically, in the case of the three-gluon vertex, the left panel of~\fig{fig:f} shows a compilation of the lattice data of this form factor in the symmetric configuration (defined as $q^2=p^2=r^2$ and $q\cd p=q\cd r=p\cd r=-q^2/2$, 
\eg 
with a $2\pi/3$ angle between each pair of momenta)~\cite{Athenodorou:2016oyh,Boucaud:2017obn}, properly normalized by dividing out the coupling [$g=2$ at $\mu=4.3$ GeV for the data set at hand, corresponding to $\alpha_s=0.32$]. Notice, in particular, the suppression of the vertex with respect to its tree-level value, as well as the sign reversal (the so-called ``zero crossing'') at small momenta,  followed by a (logarithmic) divergence at the origin. This characteristic behavior can be traced back to the delicate balance between contributions originating from gluon loops, which are ``protected'' by the corresponding gluon mass, and the ``unprotected'' logarithms coming from the ghost loops that contain (even nonperturbatively) massless ghosts~\cite{Alkofer:2008dt,Tissier:2011ey,Pelaez:2013cpa,Aguilar:2013vaa, Blum:2014gna,Eichmann:2014xya,Williams:2015cvx,Cyrol:2016tym}. 

\begin{figure*}[!t]
    \centering
    \mbox{}\hspace{-.8cm}
    \includegraphics[scale=0.6]{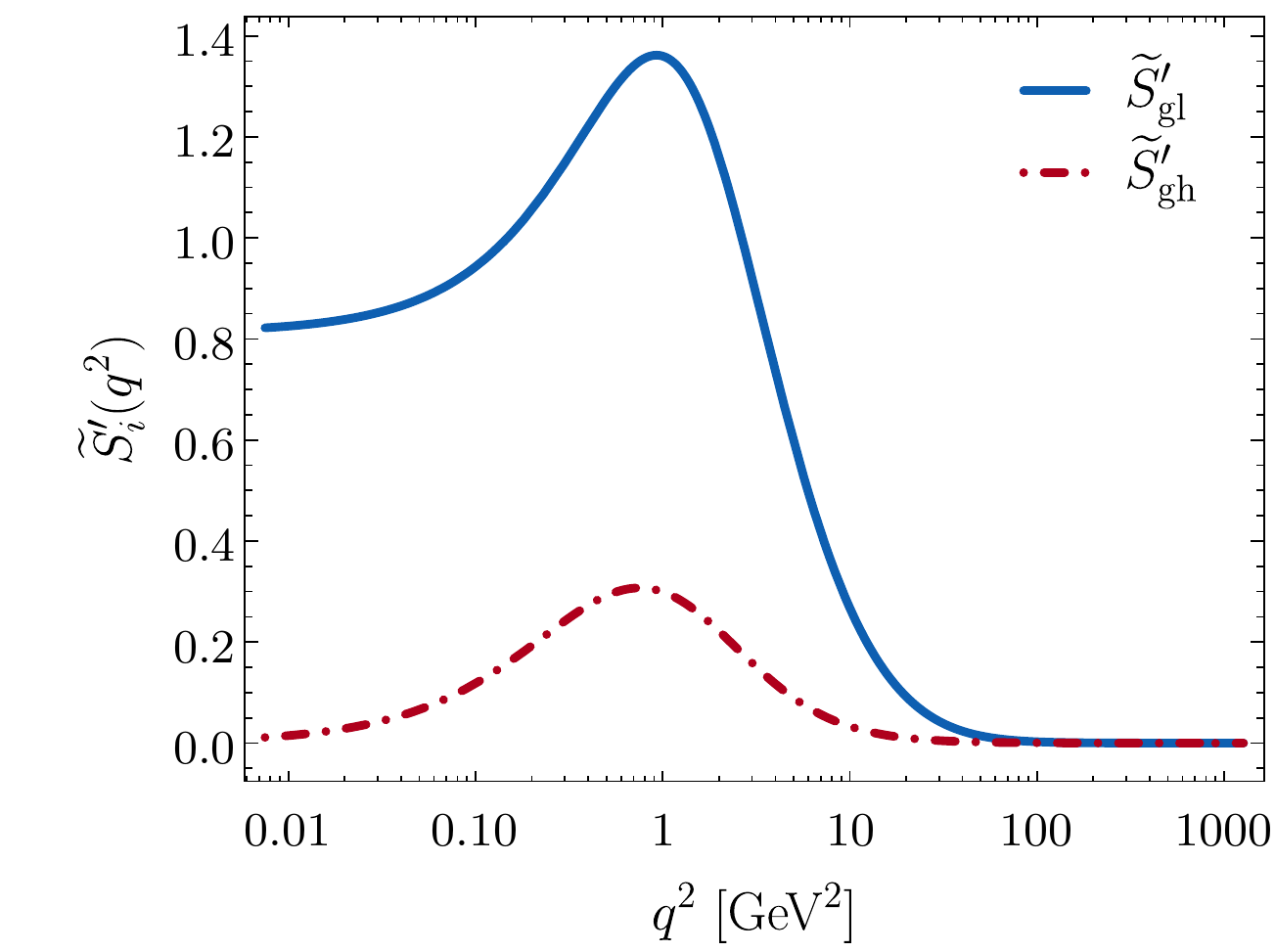}
    \hspace{.3cm}
    \includegraphics[scale=0.6]{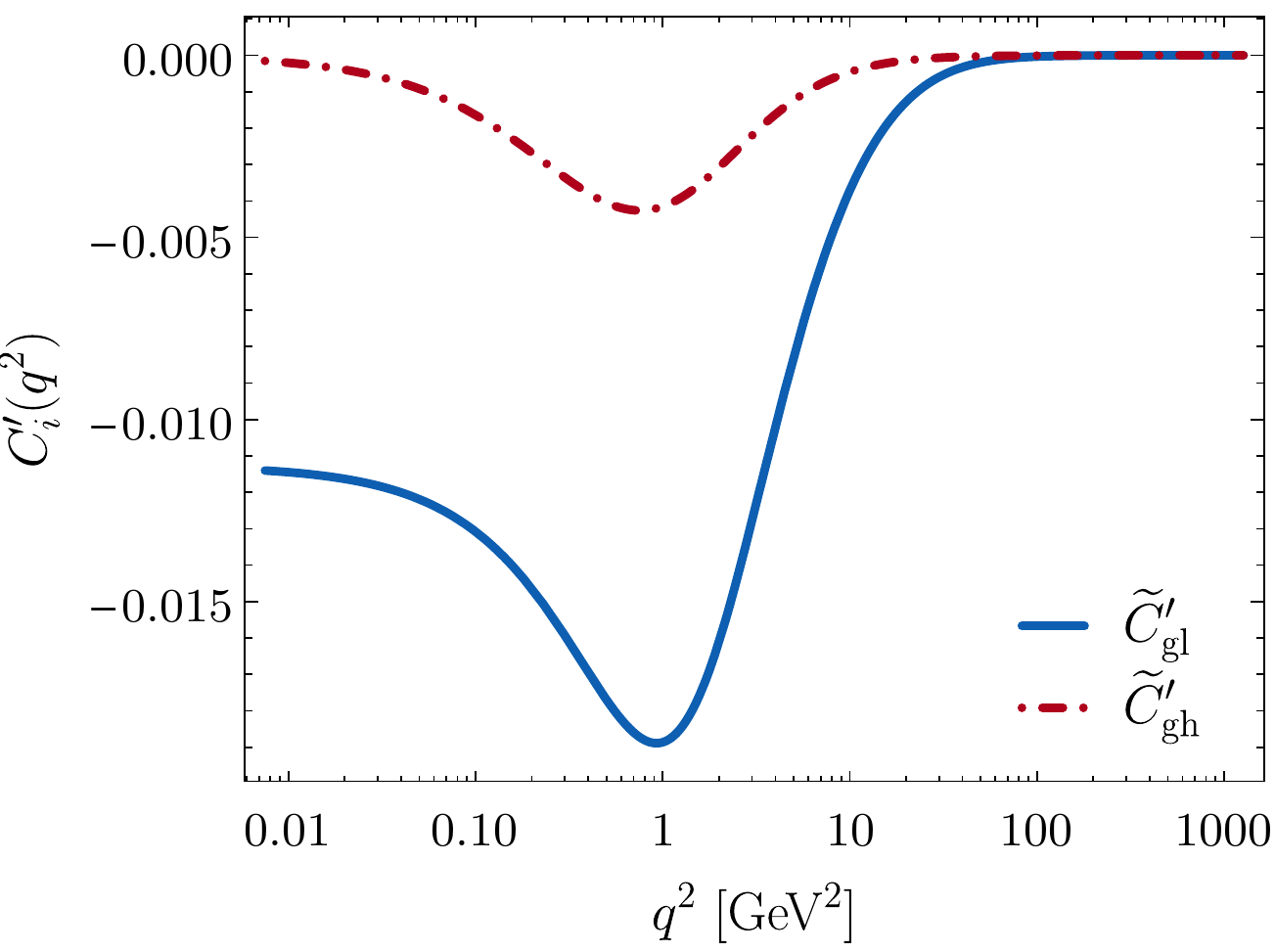}
    \caption{\label{fig:SpandCp} Unnormalized gluon and ghost solutions $\Sgl'$ and $\Sgh'$ of the BSE system~\noeq{TheSys} (left panel), and the corresponding normalized curves (right panel).}
\end{figure*}

For the ghost-gluon vertex, instead, the right panel of~\fig{fig:f} shows the numerical solution of the corresponding vertex SDE equation in the symmetric configuration within the so-called ``one-loop dressed'' approximation. The form factor is found to be equal at its tree-level value at both IR and UV values, with a characteristic peak appearing at intermediate momenta (around 0.75 GeV). The presence of this peak is in fact quite general,  appearing in 
different kinematic configurations, 
\eg the soft gluon ($q\to0$) and soft ghost ($p\to0$) limits (see respectively Fig.~6 and~7 of Ref.~\cite{Aguilar:2013xqa}). 

The (unnormalized) solutions $\Sgl^\prime$ and $\Sgh^\prime$ obtained when using these ingredients in the BSE system~\noeq{TheSys} corresponds to the eigenvalue $\aBSE=0.43$, and are shown on the left panel of~\fig{fig:SpandCp}. While it is clear that QCD dynamics is strong enough to generate 
massless poles for both vertices studied, 
the presence of a hierarchy in their relative 
``strengths" is also evident,  
as $\Sgh^\prime$ is considerably suppressed with respect to $\Sgl^\prime$ (with the latter being roughly 5 times the former at peak value).  

The common normalization constant $c$ can be determined with the procedure recently described in~\cite{Binosi:2017rwj}, that is by requiring that the normalized gluon BS amplitude give rise, when plugged into~\1eq{Cglvsmass}, to a running gluon mass that is~\n{i} monotonically decreasing and~\n{ii} vanishes in the UV. This implies~\cite{Binosi:2017rwj} $C_i^\prime=-|c|S_i^\prime$, with
\begin{align}
	|c|=\frac{\Delta^{-1}(0)}{\displaystyle\int_0^\infty\!\diff y\,\Sgl^\prime(y)},
	%\Delta^{-1}(0)/\int_0^\infty\!\diff y\,\Sgl^\prime(y),
	\label{c}
\end{align}
and, correspondingly,
\begin{align}
	\hh(q^2)=-\int_{q^2}^\infty\!\!\diff y\,\Cgl'(y).
\end{align}

\begin{figure*}[!t]
    \centering
    \mbox{}\hspace{-.8cm}
    \includegraphics[scale=0.6]{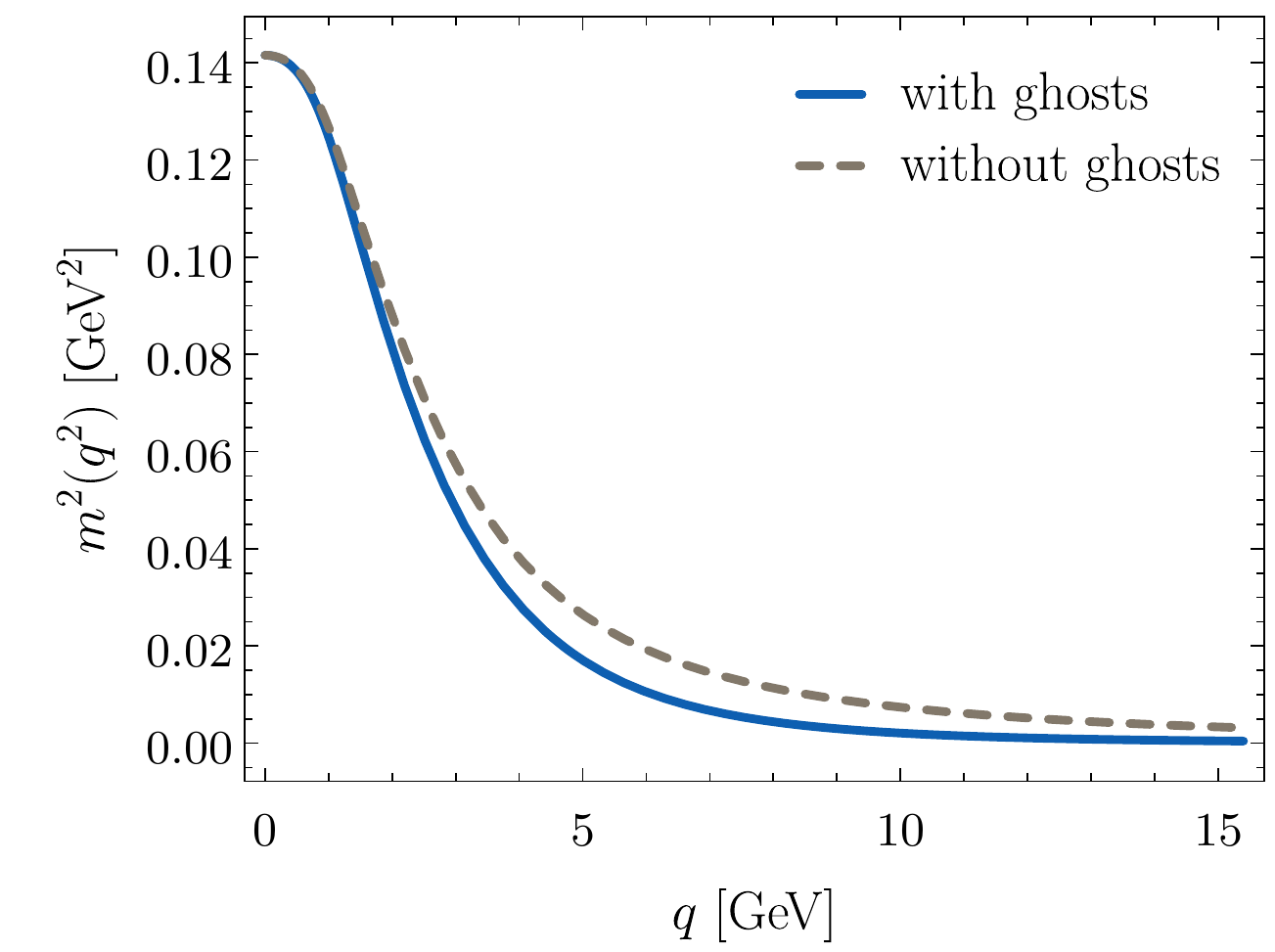}
    \caption{\label{fig:glmass} The gluon mass obtained by integrating the gluon BSE solution, compared to the one obtained in the absence of ghosts.}
\end{figure*}

The resulting gluon mass is shown in~\fig{fig:glmass}, where it is also compared to the result obtained in~\cite{Binosi:2017rwj} in the absence of ghosts, when $\aBSE=0.45$. As can be clearly appreciated, the presence of ghosts implies a faster running; indeed, one finds that the mass can be accurately fitted through the formula~\cite{Aguilar:2014tka}
\begin{align}
	m^2(q^2)=m^2(0)/[1+(q^2/m_1^2)^{1+p}], 
\end{align}
with $m_1=0.37$ GeV and $p=0.24$ as opposed to $m_1=0.36$ GeV and $p=0.1$ in the absence of ghosts.
An additional consistency check can be performed by substituting \1eq{c} into~\1eq{DSEmass},  thus obtaining a second order algebraic equation for $\alpha_s$, given by
\begin{align}
    A\alpha_s^2+B\alpha_s+C=0,
\label{quadraticmass}
\end{align}
where 
\begin{align}
    A&=\frac{3C^2_A}{32\pi^3}F(0)\hspace{-0.1cm}\int_0^\infty\hspace{-0.2cm}\diff y\, y^2{\Delta}^2(y) Y(y) \Sgl'(y),\nonumber \\
    B&=-\frac{3C_A}{8\pi} F(0)\hspace{-0.1cm}\int_0^\infty\hspace{-0.2cm}\diff y\,  \left[y^2{\Delta}^2(y) \Sgl'(y)-\frac13y^2D^2(y^2)\Sgh'(y)\right],\nonumber \\
    C&=-\int_0^\infty\hspace{-0.2cm}\diff y\,\Sgl'(y).
	\label{ABC}
\end{align}
Substituting into \1eq{ABC}
the solutions found 
for $\Sgl'(y)$ and $\Sgh'(y)$
we obtain (all values in GeV$^2$) $A=110.02$, $B=-24.25$, $C=-9.32$, yielding $\aDSE=0.43\equiv\aBSE$.

\begin{figure*}[!t]
    \centering
    \mbox{}\hspace{-.8cm}
   \includegraphics[scale=0.65]{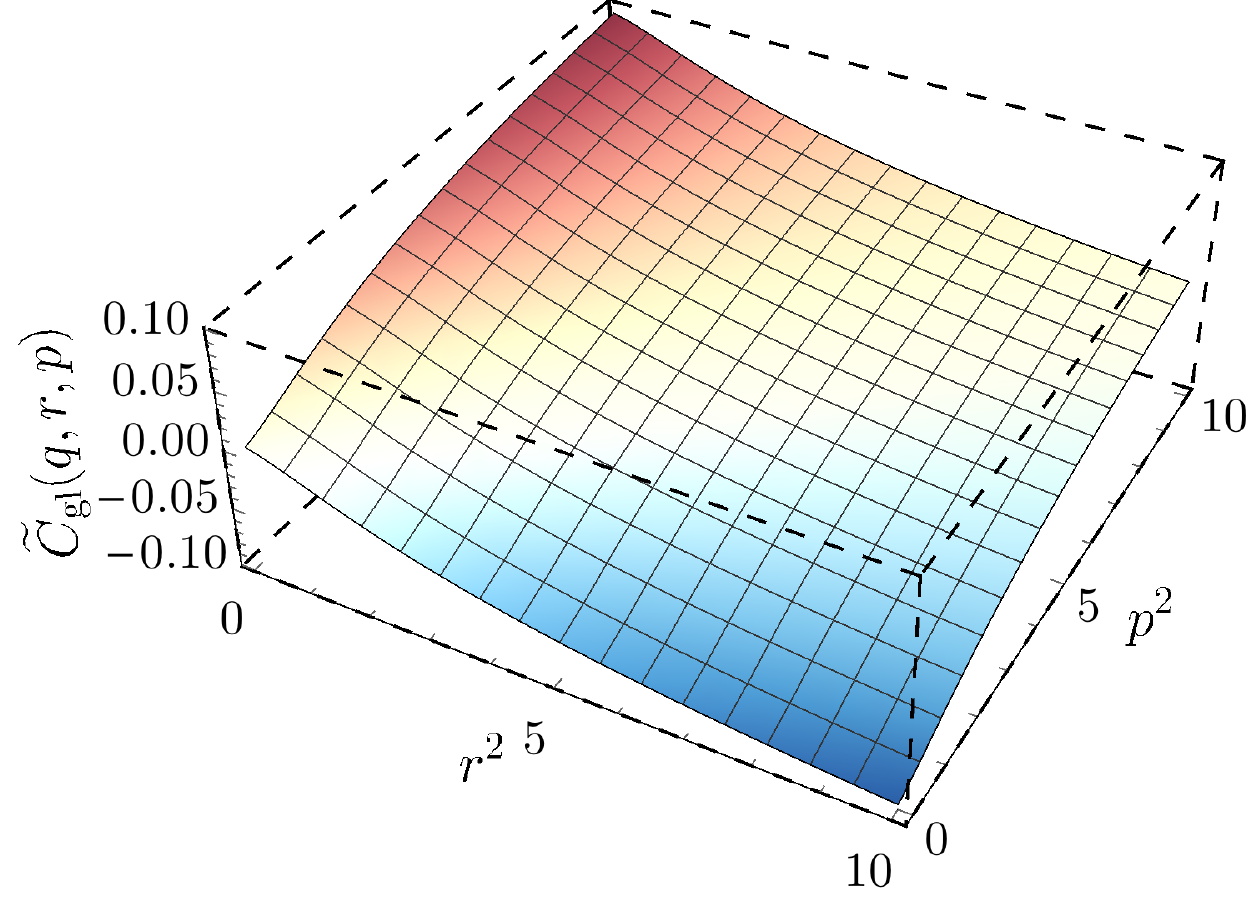}\hspace{.3cm}
    \includegraphics[scale=0.65]{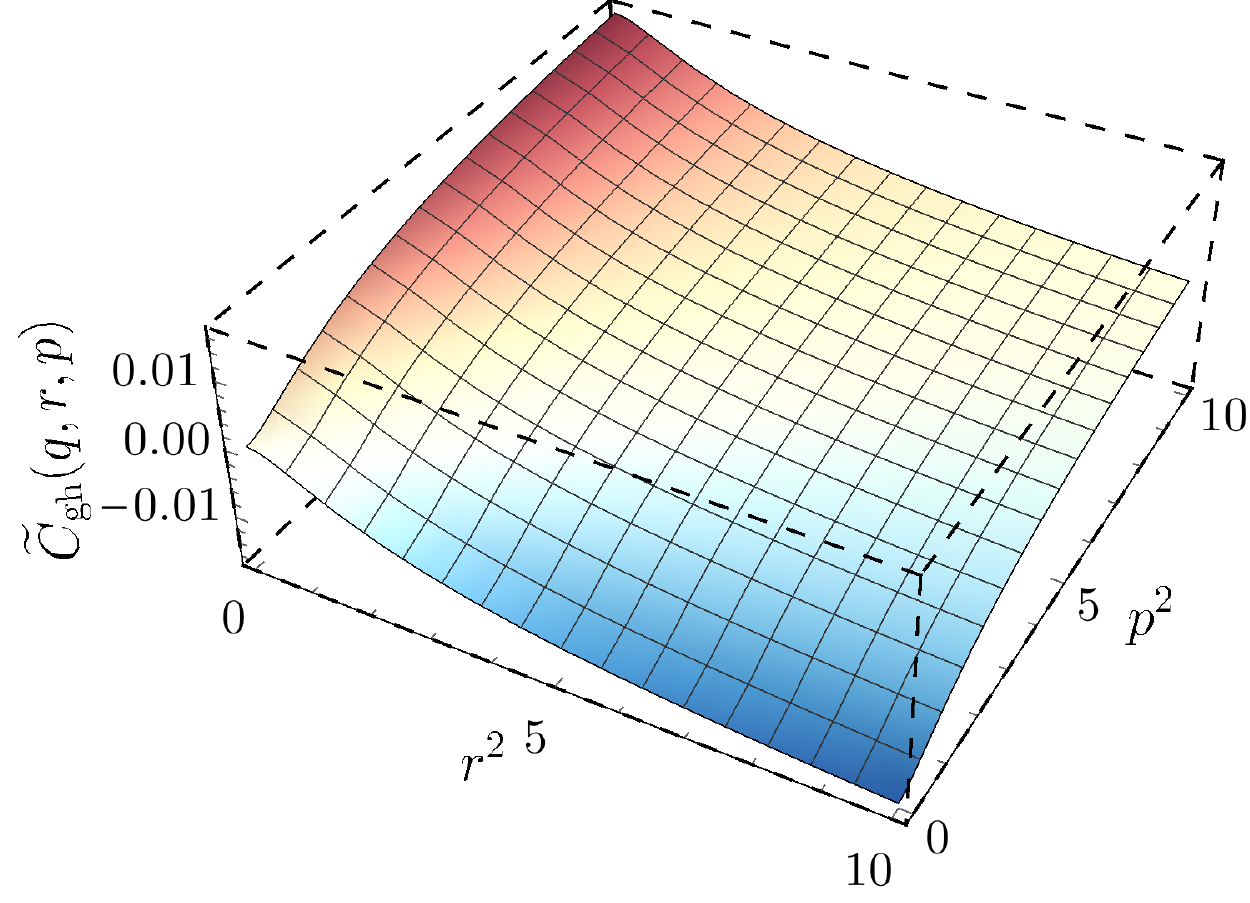}
    \caption{\label{fig:theCs} Reconstructed form factors $\Cgl$ and $\Cgh$ of the pole parts of the three-gluon and gluon-ghost vertices. The momenta $p$ and $r$ are treated as independent.}
\end{figure*}

\begin{figure*}[!t]
    \centering
    \mbox{}\hspace{-.8cm}
   \includegraphics[scale=0.6]{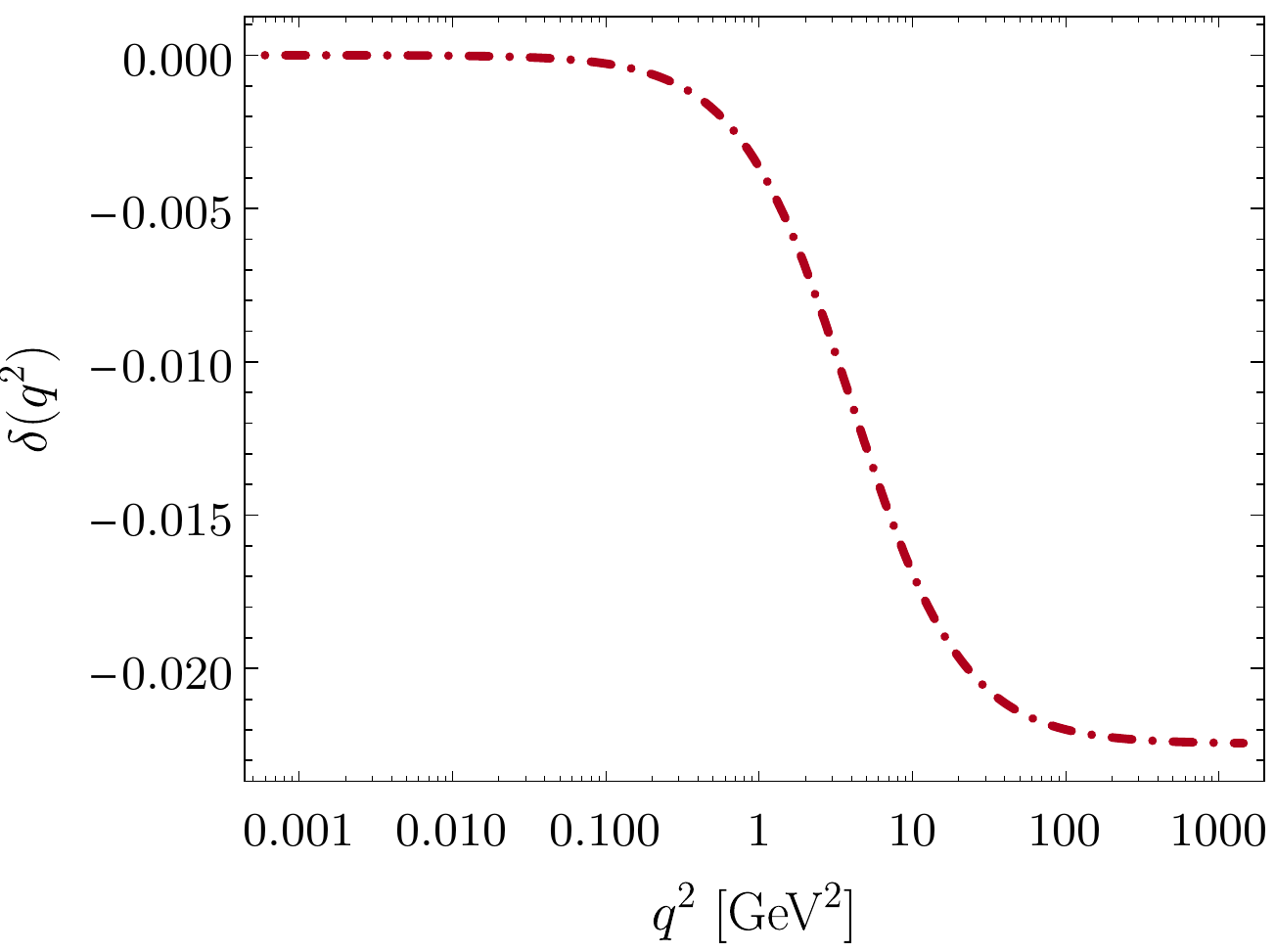}
    \caption{\label{fig:delta} The function $\delta$ measuring the relative deviation of the gluon-ghost vertex form factors from their “canonical” form, due to the presence of the pole term.}
\end{figure*}

As a final step we can fully reconstruct the form factors characterizing the three-gluon and ghost-gluon vertices pole parts, by using the results obtained so far in conjunction with~\2eqs{theCgl}{z3}. The results are shown in~\fig{fig:theCs}; notice that due to their suppression, the presence of $\Cgl$ and $\Cgh$ will not appreciably modify the no-pole parts. This can be seen also in~\fig{fig:delta} where we plot the quantity $\delta(q^2)$ introduced in~\1eq{thedelta}, which quantifies the relative deviation of the gluon-ghost vertex form factors from their ``canonical" form, due to the presence of the pole term. Such deviation saturates at the 2\% level, making the presence of poles practically undetectable from studies of three-point form factors alone.

\section{\label{conc}Conclusions}

In this work we have studied the impact of the ghost sector
on the dynamics of gluon mass generation, using the specific
framework provided by the PT-BFM formalism.
In this approach, the infrared finiteness of the gluon propagator, 
and the gluon mass connected to it, arise from the action of
massless bound state poles, which enter in the structure of the
fundamental vertices of the theory. Within this context,
our present analysis reveals that the contribution
of the poles associated with the ghost gluon vertex $\widetilde{\Gamma}_{\mu}$
are particularly suppressed with respect to those originating from the
corresponding poles of $\widetilde{\Gamma}_{\mu\alpha\beta}$.
This fact is illustrated rather clearly in~\fig{fig:theCs}, where
both vertex functions, $\Cgl(q,r,p)$ and $\Cgh(q,r,p)$, 
which accompany the corresponding poles 
and account for their relative ``strengths'', are directly compared,
for the entire range of Euclidean momenta. Evidently, whereas the
qualitative structure of both is rather similar, their relative size is
substantially different. 
Consequently, the ``gluonic'' pole contributions, $\Cgl(q,r,p)$,  
are completely decisive both for the generation and the
momentum evolution of the gluon mass.
The above result is non-trivial, in the sense that there
is no obvious {\it a-priori} argument that would imply the
observed suppression of the ghost sector. In fact,
the mere existence of solutions of 
the BSE system, let alone the observed insensitivity of the
relevant eigenvalue to the presence of $\Cgh^{\prime}(r^2)$,
may be only established once the full analysis has been carried out.

We emphasize that throughout our analysis 
we have explicitly neglected any 
possible effects stemming from poles
associated with the four-gluon vertex. In that sense, all such possible
terms have been assumed to vanish, or be numerically suppressed.
It would be clearly interesting to eventually relax this assumption
and gain some direct information of the actual size of such
contributions. Note, however, that from the technical point of view 
this task is particularly complex, mainly due to the rich tensorial
structure of this vertex~\cite{Pascual:1980yu,Binosi:2014kka,Cyrol:2014kca,Gracey:2017yfi}.
In fact, in this case the corresponding
vertex functions, $\widetilde{C}_\s{\mathrm{4gl}}(q,r,p,t)$,
depend on four rather than three kinematic variables, 
and, equivalently,
their derivatives as $q \to 0$ will depend on two instead of one,
which will vastly complicate the structure and treatment of the
would-be BSE system. 

Let us finally mention that 
an additional novel element presented in the present work
is the analysis of the behavior of $\widetilde{\Gamma}_{\mu}$
in the limit of vanishing ghost momentum, leading to the
derivation of the analogue of Taylor's theorem for the PT-BFM formalism. 
The resulting constraint relates 
one of the form factor of $\widetilde{\Gamma}_{\mu}$ with the ghost-dressing
function. In addition to its relevance for the
reconstruction of the full $\Cgh(q,r,p)$ presented here,
this particular constraint might turn useful for future lattice simulations
of the PT-BFM vertices~\cite{Binosi:2012st,Cucchieri:2012ii}, which could provide further valuable insights to this entire field of research. 

\acknowledgments 

The research of J.~P. is supported by the Spanish MEYC under grants FPA2014-53631-C2-1-P and SEV-2014-0398, and Generalitat Valenciana under grant Prometeo~II/2014/066. The work of  A.~C.~A  is supported by the National Council for Scientific and Technological Development - CNPq under the grant 305815/2015 and  by S\~ao Paulo Research Foundation - FAPESP  through the project  2017/07595-0. C.~T.~F. acknowledges the financial support from FAPESP through the fellowship 2016/11894-0. \verb|JaxoDraw|~\cite{Binosi:2003yf,Binosi:2008ig} has been used.

%\bibliography{Bibliography/bibliography}

\begin{thebibliography}{89}%
\makeatletter
\providecommand \@ifxundefined [1]{%
 \@ifx{#1\undefined}
}%
\providecommand \@ifnum [1]{%
 \ifnum #1\expandafter \@firstoftwo
 \else \expandafter \@secondoftwo
 \fi
}%
\providecommand \@ifx [1]{%
 \ifx #1\expandafter \@firstoftwo
 \else \expandafter \@secondoftwo
 \fi
}%
\providecommand \natexlab [1]{#1}%
\providecommand \enquote  [1]{``#1''}%
\providecommand \bibnamefont  [1]{#1}%
\providecommand \bibfnamefont [1]{#1}%
\providecommand \citenamefont [1]{#1}%
\providecommand \href@noop [0]{\@secondoftwo}%
\providecommand \href [0]{\begingroup \@sanitize@url \@href}%
\providecommand \@href[1]{\@@startlink{#1}\@@href}%
\providecommand \@@href[1]{\endgroup#1\@@endlink}%
\providecommand \@sanitize@url [0]{\catcode `\\12\catcode `\$12\catcode
  `\&12\catcode `\#12\catcode `\^12\catcode `\_12\catcode `\%12\relax}%
\providecommand \@@startlink[1]{}%
\providecommand \@@endlink[0]{}%
\providecommand \url  [0]{\begingroup\@sanitize@url \@url }%
\providecommand \@url [1]{\endgroup\@href {#1}{\urlprefix }}%
\providecommand \urlprefix  [0]{URL }%
\providecommand \Eprint [0]{\href }%
\providecommand \doibase [0]{http://dx.doi.org/}%
\providecommand \selectlanguage [0]{\@gobble}%
\providecommand \bibinfo  [0]{\@secondoftwo}%
\providecommand \bibfield  [0]{\@secondoftwo}%
\providecommand \translation [1]{[#1]}%
\providecommand \BibitemOpen [0]{}%
\providecommand \bibitemStop [0]{}%
\providecommand \bibitemNoStop [0]{.\EOS\space}%
\providecommand \EOS [0]{\spacefactor3000\relax}%
\providecommand \BibitemShut  [1]{\csname bibitem#1\endcsname}%
\let\auto@bib@innerbib\@empty
%</preamble>
\bibitem [{\citenamefont {Cloet}\ and\ \citenamefont
  {Roberts}(2014)}]{Cloet:2013jya}%
  \BibitemOpen
  \bibfield  {author} {\bibinfo {author} {\bibfnamefont {I.~C.}\ \bibnamefont
  {Cloet}}\ and\ \bibinfo {author} {\bibfnamefont {C.~D.}\ \bibnamefont
  {Roberts}},\ }\href {\doibase 10.1016/j.ppnp.2014.02.001} {\bibfield
  {journal} {\bibinfo  {journal} {Prog. Part. Nucl. Phys.}\ }\textbf {\bibinfo
  {volume} {77}},\ \bibinfo {pages} {1} (\bibinfo {year} {2014})},\ \Eprint
  {http://arxiv.org/abs/1310.2651} {arXiv:1310.2651 [nucl-th]} \BibitemShut
  {NoStop}%
%%CITATION = ARXIV:1310.2651;%%
\bibitem [{\citenamefont {Roberts}(2017)}]{Roberts:2016vyn}%
  \BibitemOpen
  \bibfield  {author} {\bibinfo {author} {\bibfnamefont {C.~D.}\ \bibnamefont
  {Roberts}},\ }\href {\doibase 10.1007/s00601-016-1168-z} {\bibfield
  {journal} {\bibinfo  {journal} {Few Body Syst.}\ }\textbf {\bibinfo {volume}
  {58}},\ \bibinfo {pages} {5} (\bibinfo {year} {2017})},\ \Eprint
  {http://arxiv.org/abs/1606.03909} {arXiv:1606.03909 [nucl-th]} \BibitemShut
  {NoStop}%
%%CITATION = ARXIV:1606.03909;%%
\bibitem [{\citenamefont {Roberts}\ and\ \citenamefont
  {Mezrag}(2017)}]{Roberts:2016mhh}%
  \BibitemOpen
  \bibfield  {author} {\bibinfo {author} {\bibfnamefont {C.~D.}\ \bibnamefont
  {Roberts}}\ and\ \bibinfo {author} {\bibfnamefont {C.}~\bibnamefont
  {Mezrag}},\ }\bibfield  {booktitle} {\emph {\bibinfo {booktitle}
  {{Proceedings, 12th Conference on Quark Confinement and the Hadron Spectrum
  (Confinement XII): Thessaloniki, Greece}}},\ }\href {\doibase
  10.1051/epjconf/201713701017} {\bibfield  {journal} {\bibinfo  {journal} {EPJ
  Web Conf.}\ }\textbf {\bibinfo {volume} {137}},\ \bibinfo {pages} {01017}
  (\bibinfo {year} {2017})},\ \Eprint {http://arxiv.org/abs/1611.09863}
  {arXiv:1611.09863 [nucl-th]} \BibitemShut {NoStop}%
%%CITATION = ARXIV:1611.09863;%%
\bibitem [{\citenamefont {Aguilar}\ and\ \citenamefont
  {Papavassiliou}(2006)}]{Aguilar:2006gr}%
  \BibitemOpen
  \bibfield  {author} {\bibinfo {author} {\bibfnamefont {A.~C.}\ \bibnamefont
  {Aguilar}}\ and\ \bibinfo {author} {\bibfnamefont {J.}~\bibnamefont
  {Papavassiliou}},\ }\href@noop {} {\bibfield  {journal} {\bibinfo  {journal}
  {JHEP}\ }\textbf {\bibinfo {volume} {12}},\ \bibinfo {pages} {012} (\bibinfo
  {year} {2006})},\ \Eprint {http://arxiv.org/abs/hep-ph/0610040}
  {hep-ph/0610040} \BibitemShut {NoStop}%
%%CITATION = HEP-PH/0610040;%%
\bibitem [{\citenamefont {Aguilar}\ \emph {et~al.}(2008)\citenamefont
  {Aguilar}, \citenamefont {Binosi},\ and\ \citenamefont
  {Papavassiliou}}]{Aguilar:2008xm}%
  \BibitemOpen
  \bibfield  {author} {\bibinfo {author} {\bibfnamefont {A.~C.}\ \bibnamefont
  {Aguilar}}, \bibinfo {author} {\bibfnamefont {D.}~\bibnamefont {Binosi}}, \
  and\ \bibinfo {author} {\bibfnamefont {J.}~\bibnamefont {Papavassiliou}},\
  }\href {\doibase 10.1103/PhysRevD.78.025010} {\bibfield  {journal} {\bibinfo
  {journal} {Phys. Rev.}\ }\textbf {\bibinfo {volume} {D78}},\ \bibinfo {pages}
  {025010} (\bibinfo {year} {2008})},\ \Eprint {http://arxiv.org/abs/0802.1870}
  {arXiv:0802.1870 [hep-ph]} \BibitemShut {NoStop}%
\bibitem [{\citenamefont {Aguilar}\ \emph {et~al.}(2012)\citenamefont
  {Aguilar}, \citenamefont {Ibanez}, \citenamefont {Mathieu},\ and\
  \citenamefont {Papavassiliou}}]{Aguilar:2011xe}%
  \BibitemOpen
  \bibfield  {author} {\bibinfo {author} {\bibfnamefont {A.}~\bibnamefont
  {Aguilar}}, \bibinfo {author} {\bibfnamefont {D.}~\bibnamefont {Ibanez}},
  \bibinfo {author} {\bibfnamefont {V.}~\bibnamefont {Mathieu}}, \ and\
  \bibinfo {author} {\bibfnamefont {J.}~\bibnamefont {Papavassiliou}},\ }\href
  {\doibase 10.1103/PhysRevD.85.014018} {\bibfield  {journal} {\bibinfo
  {journal} {Phys.~Rev.~}\ }\textbf {\bibinfo {volume} {D85}},\ \bibinfo
  {pages} {014018} (\bibinfo {year} {2012})},\ \Eprint
  {http://arxiv.org/abs/1110.2633} {arXiv:1110.2633 [hep-ph]} \BibitemShut
  {NoStop}%
%%CITATION = ARXIV:1110.2633;%%
\bibitem [{\citenamefont {Iba{\~n}ez}\ and\ \citenamefont
  {Papavassiliou}(2013)}]{Ibanez:2012zk}%
  \BibitemOpen
  \bibfield  {author} {\bibinfo {author} {\bibfnamefont {D.}~\bibnamefont
  {Iba{\~n}ez}}\ and\ \bibinfo {author} {\bibfnamefont {J.}~\bibnamefont
  {Papavassiliou}},\ }\href {\doibase 10.1103/PhysRevD.87.034008} {\bibfield
  {journal} {\bibinfo  {journal} {Phys.~Rev.~}\ }\textbf {\bibinfo {volume}
  {D87}},\ \bibinfo {pages} {034008} (\bibinfo {year} {2013})},\ \Eprint
  {http://arxiv.org/abs/1211.5314} {arXiv:1211.5314 [hep-ph]} \BibitemShut
  {NoStop}%
%%CITATION = ARXIV:1211.5314;%%
\bibitem [{\citenamefont {Aguilar}\ \emph {et~al.}(2016)\citenamefont
  {Aguilar}, \citenamefont {Binosi}, \citenamefont {Figueiredo},\ and\
  \citenamefont {Papavassiliou}}]{Aguilar:2016vin}%
  \BibitemOpen
  \bibfield  {author} {\bibinfo {author} {\bibfnamefont {A.~C.}\ \bibnamefont
  {Aguilar}}, \bibinfo {author} {\bibfnamefont {D.}~\bibnamefont {Binosi}},
  \bibinfo {author} {\bibfnamefont {C.~T.}\ \bibnamefont {Figueiredo}}, \ and\
  \bibinfo {author} {\bibfnamefont {J.}~\bibnamefont {Papavassiliou}},\ }\href
  {\doibase 10.1103/PhysRevD.94.045002} {\bibfield  {journal} {\bibinfo
  {journal} {Phys. Rev.}\ }\textbf {\bibinfo {volume} {D94}},\ \bibinfo {pages}
  {045002} (\bibinfo {year} {2016})},\ \Eprint
  {http://arxiv.org/abs/1604.08456} {arXiv:1604.08456 [hep-ph]} \BibitemShut
  {NoStop}%
%%CITATION = ARXIV:1604.08456;%%
\bibitem [{\citenamefont {Cornwall}(1982)}]{Cornwall:1981zr}%
  \BibitemOpen
  \bibfield  {author} {\bibinfo {author} {\bibfnamefont {J.~M.}\ \bibnamefont
  {Cornwall}},\ }\href@noop {} {\bibfield  {journal} {\bibinfo  {journal}
  {Phys. Rev.}\ }\textbf {\bibinfo {volume} {D26}},\ \bibinfo {pages} {1453}
  (\bibinfo {year} {1982})}\BibitemShut {NoStop}%
%%CITATION = PHRVA,D26,1453;%%
\bibitem [{\citenamefont {Cucchieri}\ and\ \citenamefont
  {Mendes}(2007)}]{Cucchieri:2007md}%
  \BibitemOpen
  \bibfield  {author} {\bibinfo {author} {\bibfnamefont {A.}~\bibnamefont
  {Cucchieri}}\ and\ \bibinfo {author} {\bibfnamefont {T.}~\bibnamefont
  {Mendes}},\ }\href@noop {} {\bibfield  {journal} {\bibinfo  {journal} {PoS}\
  }\textbf {\bibinfo {volume} {LAT2007}},\ \bibinfo {pages} {297} (\bibinfo
  {year} {2007})},\ \Eprint {http://arxiv.org/abs/0710.0412} {arXiv:0710.0412
  [hep-lat]} \BibitemShut {NoStop}%
%%CITATION = ARXIV:0710.0412;%%
\bibitem [{\citenamefont {Cucchieri}\ and\ \citenamefont
  {Mendes}(2008)}]{Cucchieri:2007rg}%
  \BibitemOpen
  \bibfield  {author} {\bibinfo {author} {\bibfnamefont {A.}~\bibnamefont
  {Cucchieri}}\ and\ \bibinfo {author} {\bibfnamefont {T.}~\bibnamefont
  {Mendes}},\ }\href {\doibase 10.1103/PhysRevLett.100.241601} {\bibfield
  {journal} {\bibinfo  {journal} {Phys.~Rev.~Lett.}\ }\textbf {\bibinfo
  {volume} {100}},\ \bibinfo {pages} {241601} (\bibinfo {year} {2008})},\
  \Eprint {http://arxiv.org/abs/0712.3517} {arXiv:0712.3517 [hep-lat]}
  \BibitemShut {NoStop}%
%%CITATION = ARXIV:0712.3517;%%
\bibitem [{\citenamefont {Cucchieri}\ and\ \citenamefont
  {Mendes}(2010)}]{Cucchieri:2009zt}%
  \BibitemOpen
  \bibfield  {author} {\bibinfo {author} {\bibfnamefont {A.}~\bibnamefont
  {Cucchieri}}\ and\ \bibinfo {author} {\bibfnamefont {T.}~\bibnamefont
  {Mendes}},\ }\href {\doibase 10.1103/PhysRevD.81.016005} {\bibfield
  {journal} {\bibinfo  {journal} {Phys.~Rev.~}\ }\textbf {\bibinfo {volume}
  {D81}},\ \bibinfo {pages} {016005} (\bibinfo {year} {2010})},\ \Eprint
  {http://arxiv.org/abs/0904.4033} {arXiv:0904.4033 [hep-lat]} \BibitemShut
  {NoStop}%
%%CITATION = ARXIV:0904.4033;%%
\bibitem [{\citenamefont {Bowman}\ \emph {et~al.}(2007)\citenamefont {Bowman}
  \emph {et~al.}}]{Bowman:2007du}%
  \BibitemOpen
  \bibfield  {author} {\bibinfo {author} {\bibfnamefont {P.~O.}\ \bibnamefont
  {Bowman}} \emph {et~al.},\ }\href@noop {} {\bibfield  {journal} {\bibinfo
  {journal} {Phys. Rev.}\ }\textbf {\bibinfo {volume} {D76}},\ \bibinfo {pages}
  {094505} (\bibinfo {year} {2007})},\ \Eprint
  {http://arxiv.org/abs/hep-lat/0703022} {hep-lat/0703022} \BibitemShut
  {NoStop}%
%%CITATION = HEP-LAT/0703022;%%
\bibitem [{\citenamefont {Bogolubsky}\ \emph {et~al.}(2009)\citenamefont
  {Bogolubsky}, \citenamefont {Ilgenfritz}, \citenamefont {Muller-Preussker},\
  and\ \citenamefont {Sternbeck}}]{Bogolubsky:2009dc}%
  \BibitemOpen
  \bibfield  {author} {\bibinfo {author} {\bibfnamefont {I.}~\bibnamefont
  {Bogolubsky}}, \bibinfo {author} {\bibfnamefont {E.}~\bibnamefont
  {Ilgenfritz}}, \bibinfo {author} {\bibfnamefont {M.}~\bibnamefont
  {Muller-Preussker}}, \ and\ \bibinfo {author} {\bibfnamefont
  {A.}~\bibnamefont {Sternbeck}},\ }\href {\doibase
  10.1016/j.physletb.2009.04.076} {\bibfield  {journal} {\bibinfo  {journal}
  {Phys. Lett.}\ }\textbf {\bibinfo {volume} {B676}},\ \bibinfo {pages} {69}
  (\bibinfo {year} {2009})},\ \Eprint {http://arxiv.org/abs/0901.0736}
  {arXiv:0901.0736 [hep-lat]} \BibitemShut {NoStop}%
%%CITATION = ARXIV:0901.0736;%%
\bibitem [{\citenamefont {Oliveira}\ and\ \citenamefont
  {Silva}(2009)}]{Oliveira:2009eh}%
  \BibitemOpen
  \bibfield  {author} {\bibinfo {author} {\bibfnamefont {O.}~\bibnamefont
  {Oliveira}}\ and\ \bibinfo {author} {\bibfnamefont {P.}~\bibnamefont
  {Silva}},\ }\href@noop {} {\bibfield  {journal} {\bibinfo  {journal} {PoS}\
  }\textbf {\bibinfo {volume} {LAT2009}},\ \bibinfo {pages} {226} (\bibinfo
  {year} {2009})},\ \Eprint {http://arxiv.org/abs/0910.2897} {arXiv:0910.2897
  [hep-lat]} \BibitemShut {NoStop}%
%%CITATION = ARXIV:0910.2897;%%
\bibitem [{\citenamefont {Ayala}\ \emph {et~al.}(2012)\citenamefont {Ayala},
  \citenamefont {Bashir}, \citenamefont {Binosi}, \citenamefont
  {Cristoforetti},\ and\ \citenamefont {Rodriguez-Quintero}}]{Ayala:2012pb}%
  \BibitemOpen
  \bibfield  {author} {\bibinfo {author} {\bibfnamefont {A.}~\bibnamefont
  {Ayala}}, \bibinfo {author} {\bibfnamefont {A.}~\bibnamefont {Bashir}},
  \bibinfo {author} {\bibfnamefont {D.}~\bibnamefont {Binosi}}, \bibinfo
  {author} {\bibfnamefont {M.}~\bibnamefont {Cristoforetti}}, \ and\ \bibinfo
  {author} {\bibfnamefont {J.}~\bibnamefont {Rodriguez-Quintero}},\ }\href
  {\doibase 10.1103/PhysRevD.86.074512} {\bibfield  {journal} {\bibinfo
  {journal} {Phys. Rev.}\ }\textbf {\bibinfo {volume} {D86}},\ \bibinfo {pages}
  {074512} (\bibinfo {year} {2012})},\ \Eprint {http://arxiv.org/abs/1208.0795}
  {arXiv:1208.0795 [hep-ph]} \BibitemShut {NoStop}%
%%CITATION = ARXIV:1208.0795;%%
\bibitem [{\citenamefont {Bicudo}\ \emph {et~al.}(2015)\citenamefont {Bicudo},
  \citenamefont {Binosi}, \citenamefont {Cardoso}, \citenamefont {Oliveira},\
  and\ \citenamefont {Silva}}]{Bicudo:2015rma}%
  \BibitemOpen
  \bibfield  {author} {\bibinfo {author} {\bibfnamefont {P.}~\bibnamefont
  {Bicudo}}, \bibinfo {author} {\bibfnamefont {D.}~\bibnamefont {Binosi}},
  \bibinfo {author} {\bibfnamefont {N.}~\bibnamefont {Cardoso}}, \bibinfo
  {author} {\bibfnamefont {O.}~\bibnamefont {Oliveira}}, \ and\ \bibinfo
  {author} {\bibfnamefont {P.~J.}\ \bibnamefont {Silva}},\ }\href {\doibase
  10.1103/PhysRevD.92.114514} {\bibfield  {journal} {\bibinfo  {journal} {Phys.
  Rev.}\ }\textbf {\bibinfo {volume} {D92}},\ \bibinfo {pages} {114514}
  (\bibinfo {year} {2015})},\ \Eprint {http://arxiv.org/abs/1505.05897}
  {arXiv:1505.05897 [hep-lat]} \BibitemShut {NoStop}%
%%CITATION = ARXIV:1505.05897;%%
\bibitem [{\citenamefont {Lavelle}(1991)}]{Lavelle:1991ve}%
  \BibitemOpen
  \bibfield  {author} {\bibinfo {author} {\bibfnamefont {M.}~\bibnamefont
  {Lavelle}},\ }\href@noop {} {\bibfield  {journal} {\bibinfo  {journal} {Phys.
  Rev.}\ }\textbf {\bibinfo {volume} {D44}},\ \bibinfo {pages} {26} (\bibinfo
  {year} {1991})}\BibitemShut {NoStop}%
%%CITATION = PHRVA,D44,26;%%
\bibitem [{\citenamefont {Halzen}\ \emph {et~al.}(1993)\citenamefont {Halzen},
  \citenamefont {Krein},\ and\ \citenamefont {Natale}}]{Halzen:1992vd}%
  \BibitemOpen
  \bibfield  {author} {\bibinfo {author} {\bibfnamefont {F.}~\bibnamefont
  {Halzen}}, \bibinfo {author} {\bibfnamefont {G.~I.}\ \bibnamefont {Krein}}, \
  and\ \bibinfo {author} {\bibfnamefont {A.~A.}\ \bibnamefont {Natale}},\
  }\href@noop {} {\bibfield  {journal} {\bibinfo  {journal} {Phys. Rev.}\
  }\textbf {\bibinfo {volume} {D47}},\ \bibinfo {pages} {295} (\bibinfo {year}
  {1993})}\BibitemShut {NoStop}%
%%CITATION = PHRVA,D47,295;%%
\bibitem [{\citenamefont {Philipsen}(2002)}]{Philipsen:2001ip}%
  \BibitemOpen
  \bibfield  {author} {\bibinfo {author} {\bibfnamefont {O.}~\bibnamefont
  {Philipsen}},\ }\href {\doibase 10.1016/S0550-3213(02)00089-5} {\bibfield
  {journal} {\bibinfo  {journal} {Nucl. Phys.}\ }\textbf {\bibinfo {volume}
  {B628}},\ \bibinfo {pages} {167} (\bibinfo {year} {2002})},\ \Eprint
  {http://arxiv.org/abs/hep-lat/0112047} {arXiv:hep-lat/0112047 [hep-lat]}
  \BibitemShut {NoStop}%
%%CITATION = HEP-LAT/0112047;%%
\bibitem [{\citenamefont {Szczepaniak}\ and\ \citenamefont
  {Swanson}(2002)}]{Szczepaniak:2001rg}%
  \BibitemOpen
  \bibfield  {author} {\bibinfo {author} {\bibfnamefont {A.~P.}\ \bibnamefont
  {Szczepaniak}}\ and\ \bibinfo {author} {\bibfnamefont {E.~S.}\ \bibnamefont
  {Swanson}},\ }\href {\doibase 10.1103/PhysRevD.65.025012} {\bibfield
  {journal} {\bibinfo  {journal} {Phys. Rev.}\ }\textbf {\bibinfo {volume}
  {D65}},\ \bibinfo {pages} {025012} (\bibinfo {year} {2002})},\ \Eprint
  {http://arxiv.org/abs/hep-ph/0107078} {arXiv:hep-ph/0107078 [hep-ph]}
  \BibitemShut {NoStop}%
%%CITATION = HEP-PH/0107078;%%
\bibitem [{\citenamefont {Aguilar}\ and\ \citenamefont
  {Natale}(2004)}]{Aguilar:2004sw}%
  \BibitemOpen
  \bibfield  {author} {\bibinfo {author} {\bibfnamefont {A.~C.}\ \bibnamefont
  {Aguilar}}\ and\ \bibinfo {author} {\bibfnamefont {A.~A.}\ \bibnamefont
  {Natale}},\ }\href@noop {} {\bibfield  {journal} {\bibinfo  {journal} {JHEP}\
  }\textbf {\bibinfo {volume} {08}},\ \bibinfo {pages} {057} (\bibinfo {year}
  {2004})},\ \Eprint {http://arxiv.org/abs/hep-ph/0408254} {hep-ph/0408254}
  \BibitemShut {NoStop}%
%%CITATION = HEP-PH/0408254;%%
\bibitem [{\citenamefont {Kondo}(2006)}]{Kondo:2006ih}%
  \BibitemOpen
  \bibfield  {author} {\bibinfo {author} {\bibfnamefont {K.-I.}\ \bibnamefont
  {Kondo}},\ }\href@noop {} {\bibfield  {journal} {\bibinfo  {journal} {Phys.
  Rev.}\ }\textbf {\bibinfo {volume} {D74}},\ \bibinfo {pages} {125003}
  (\bibinfo {year} {2006})},\ \Eprint {http://arxiv.org/abs/hep-th/0609166}
  {hep-th/0609166} \BibitemShut {NoStop}%
%%CITATION = HEP-TH/0609166;%%
\bibitem [{\citenamefont {Braun}\ \emph {et~al.}(2010)\citenamefont {Braun},
  \citenamefont {Gies},\ and\ \citenamefont {Pawlowski}}]{Braun:2007bx}%
  \BibitemOpen
  \bibfield  {author} {\bibinfo {author} {\bibfnamefont {J.}~\bibnamefont
  {Braun}}, \bibinfo {author} {\bibfnamefont {H.}~\bibnamefont {Gies}}, \ and\
  \bibinfo {author} {\bibfnamefont {J.~M.}\ \bibnamefont {Pawlowski}},\ }\href
  {\doibase 10.1016/j.physletb.2010.01.009} {\bibfield  {journal} {\bibinfo
  {journal} {Phys.~Lett.~}\ }\textbf {\bibinfo {volume} {B684}},\ \bibinfo
  {pages} {262} (\bibinfo {year} {2010})},\ \Eprint
  {http://arxiv.org/abs/0708.2413} {arXiv:0708.2413 [hep-th]} \BibitemShut
  {NoStop}%
%%CITATION = ARXIV:0708.2413;%%
\bibitem [{\citenamefont {Epple}\ \emph {et~al.}(2008)\citenamefont {Epple},
  \citenamefont {Reinhardt}, \citenamefont {Schleifenbaum},\ and\ \citenamefont
  {Szczepaniak}}]{Epple:2007ut}%
  \BibitemOpen
  \bibfield  {author} {\bibinfo {author} {\bibfnamefont {D.}~\bibnamefont
  {Epple}}, \bibinfo {author} {\bibfnamefont {H.}~\bibnamefont {Reinhardt}},
  \bibinfo {author} {\bibfnamefont {W.}~\bibnamefont {Schleifenbaum}}, \ and\
  \bibinfo {author} {\bibfnamefont {A.}~\bibnamefont {Szczepaniak}},\ }\href
  {\doibase 10.1103/PhysRevD.77.085007} {\bibfield  {journal} {\bibinfo
  {journal} {Phys. Rev.}\ }\textbf {\bibinfo {volume} {D77}},\ \bibinfo {pages}
  {085007} (\bibinfo {year} {2008})},\ \Eprint {http://arxiv.org/abs/0712.3694}
  {arXiv:0712.3694 [hep-th]} \BibitemShut {NoStop}%
%%CITATION = ARXIV:0712.3694;%%
\bibitem [{\citenamefont {Boucaud}\ \emph {et~al.}(2008)\citenamefont {Boucaud}
  \emph {et~al.}}]{Boucaud:2008ky}%
  \BibitemOpen
  \bibfield  {author} {\bibinfo {author} {\bibfnamefont {P.}~\bibnamefont
  {Boucaud}} \emph {et~al.},\ }\href {\doibase 10.1088/1126-6708/2008/06/099}
  {\bibfield  {journal} {\bibinfo  {journal} {JHEP}\ }\textbf {\bibinfo
  {volume} {06}},\ \bibinfo {pages} {099} (\bibinfo {year} {2008})},\ \Eprint
  {http://arxiv.org/abs/0803.2161} {arXiv:0803.2161 [hep-ph]} \BibitemShut
  {NoStop}%
%%CITATION = 0803.2161;%%
\bibitem [{\citenamefont {Dudal}\ \emph {et~al.}(2008)\citenamefont {Dudal},
  \citenamefont {Gracey}, \citenamefont {Sorella}, \citenamefont
  {Vandersickel},\ and\ \citenamefont {Verschelde}}]{Dudal:2008sp}%
  \BibitemOpen
  \bibfield  {author} {\bibinfo {author} {\bibfnamefont {D.}~\bibnamefont
  {Dudal}}, \bibinfo {author} {\bibfnamefont {J.~A.}\ \bibnamefont {Gracey}},
  \bibinfo {author} {\bibfnamefont {S.~P.}\ \bibnamefont {Sorella}}, \bibinfo
  {author} {\bibfnamefont {N.}~\bibnamefont {Vandersickel}}, \ and\ \bibinfo
  {author} {\bibfnamefont {H.}~\bibnamefont {Verschelde}},\ }\href {\doibase
  10.1103/PhysRevD.78.065047} {\bibfield  {journal} {\bibinfo  {journal} {Phys.
  Rev.}\ }\textbf {\bibinfo {volume} {D78}},\ \bibinfo {pages} {065047}
  (\bibinfo {year} {2008})},\ \Eprint {http://arxiv.org/abs/0806.4348}
  {arXiv:0806.4348 [hep-th]} \BibitemShut {NoStop}%
%%CITATION = 0806.4348;%%
\bibitem [{\citenamefont {Fischer}\ \emph {et~al.}(2009)\citenamefont
  {Fischer}, \citenamefont {Maas},\ and\ \citenamefont
  {Pawlowski}}]{Fischer:2008uz}%
  \BibitemOpen
  \bibfield  {author} {\bibinfo {author} {\bibfnamefont {C.~S.}\ \bibnamefont
  {Fischer}}, \bibinfo {author} {\bibfnamefont {A.}~\bibnamefont {Maas}}, \
  and\ \bibinfo {author} {\bibfnamefont {J.~M.}\ \bibnamefont {Pawlowski}},\
  }\href {\doibase 10.1016/j.aop.2009.07.009} {\bibfield  {journal} {\bibinfo
  {journal} {Annals Phys.}\ }\textbf {\bibinfo {volume} {324}},\ \bibinfo
  {pages} {2408} (\bibinfo {year} {2009})},\ \Eprint
  {http://arxiv.org/abs/0810.1987} {arXiv:0810.1987 [hep-ph]} \BibitemShut
  {NoStop}%
%%CITATION = ARXIV:0810.1987;%%
\bibitem [{\citenamefont {Aguilar}\ \emph {et~al.}(2009)\citenamefont
  {Aguilar}, \citenamefont {Binosi}, \citenamefont {Papavassiliou},\ and\
  \citenamefont {Rodriguez-Quintero}}]{Aguilar:2009nf}%
  \BibitemOpen
  \bibfield  {author} {\bibinfo {author} {\bibfnamefont {A.~C.}\ \bibnamefont
  {Aguilar}}, \bibinfo {author} {\bibfnamefont {D.}~\bibnamefont {Binosi}},
  \bibinfo {author} {\bibfnamefont {J.}~\bibnamefont {Papavassiliou}}, \ and\
  \bibinfo {author} {\bibfnamefont {J.}~\bibnamefont {Rodriguez-Quintero}},\
  }\href {\doibase 10.1103/PhysRevD.80.085018} {\bibfield  {journal} {\bibinfo
  {journal} {Phys. Rev.}\ }\textbf {\bibinfo {volume} {D80}},\ \bibinfo {pages}
  {085018} (\bibinfo {year} {2009})},\ \Eprint {http://arxiv.org/abs/0906.2633}
  {arXiv:0906.2633 [hep-ph]} \BibitemShut {NoStop}%
\bibitem [{\citenamefont
  {Rodriguez-Quintero}(2011)}]{RodriguezQuintero:2010wy}%
  \BibitemOpen
  \bibfield  {author} {\bibinfo {author} {\bibfnamefont {J.}~\bibnamefont
  {Rodriguez-Quintero}},\ }\href {\doibase 10.1007/JHEP01(2011)105} {\bibfield
  {journal} {\bibinfo  {journal} {JHEP}\ }\textbf {\bibinfo {volume} {1101}},\
  \bibinfo {pages} {105} (\bibinfo {year} {2011})},\ \Eprint
  {http://arxiv.org/abs/1005.4598} {arXiv:1005.4598 [hep-ph]} \BibitemShut
  {NoStop}%
%%CITATION = ARXIV:1005.4598;%%
\bibitem [{\citenamefont {Campagnari}\ and\ \citenamefont
  {Reinhardt}(2010)}]{Campagnari:2010wc}%
  \BibitemOpen
  \bibfield  {author} {\bibinfo {author} {\bibfnamefont {D.~R.}\ \bibnamefont
  {Campagnari}}\ and\ \bibinfo {author} {\bibfnamefont {H.}~\bibnamefont
  {Reinhardt}},\ }\href {\doibase 10.1103/PhysRevD.82.105021} {\bibfield
  {journal} {\bibinfo  {journal} {Phys. Rev.}\ }\textbf {\bibinfo {volume}
  {D82}},\ \bibinfo {pages} {105021} (\bibinfo {year} {2010})},\ \Eprint
  {http://arxiv.org/abs/1009.4599} {arXiv:1009.4599 [hep-th]} \BibitemShut
  {NoStop}%
%%CITATION = ARXIV:1009.4599;%%
\bibitem [{\citenamefont {Tissier}\ and\ \citenamefont
  {Wschebor}(2010)}]{Tissier:2010ts}%
  \BibitemOpen
  \bibfield  {author} {\bibinfo {author} {\bibfnamefont {M.}~\bibnamefont
  {Tissier}}\ and\ \bibinfo {author} {\bibfnamefont {N.}~\bibnamefont
  {Wschebor}},\ }\href {\doibase 10.1103/PhysRevD.82.101701} {\bibfield
  {journal} {\bibinfo  {journal} {Phys.~Rev.~}\ }\textbf {\bibinfo {volume}
  {D82}},\ \bibinfo {pages} {101701} (\bibinfo {year} {2010})},\ \Eprint
  {http://arxiv.org/abs/1004.1607} {arXiv:1004.1607 [hep-ph]} \BibitemShut
  {NoStop}%
%%CITATION = ARXIV:1004.1607;%%
\bibitem [{\citenamefont {Kondo}(2010)}]{Kondo:2010ts}%
  \BibitemOpen
  \bibfield  {author} {\bibinfo {author} {\bibfnamefont {K.-I.}\ \bibnamefont
  {Kondo}},\ }\href {\doibase 10.1103/PhysRevD.82.065024} {\bibfield  {journal}
  {\bibinfo  {journal} {Phys. Rev.}\ }\textbf {\bibinfo {volume} {D82}},\
  \bibinfo {pages} {065024} (\bibinfo {year} {2010})},\ \Eprint
  {http://arxiv.org/abs/1005.0314} {arXiv:1005.0314 [hep-th]} \BibitemShut
  {NoStop}%
%%CITATION = ARXIV:1005.0314;%%
\bibitem [{\citenamefont {Pennington}\ and\ \citenamefont
  {Wilson}(2011)}]{Pennington:2011xs}%
  \BibitemOpen
  \bibfield  {author} {\bibinfo {author} {\bibfnamefont {M.}~\bibnamefont
  {Pennington}}\ and\ \bibinfo {author} {\bibfnamefont {D.}~\bibnamefont
  {Wilson}},\ }\href {\doibase 10.1103/PhysRevD.84.094028,
  10.1103/PhysRevD.84.119901} {\bibfield  {journal} {\bibinfo  {journal} {Phys.
  Rev.}\ }\textbf {\bibinfo {volume} {D84}},\ \bibinfo {pages} {119901}
  (\bibinfo {year} {2011})},\ \Eprint {http://arxiv.org/abs/1109.2117}
  {arXiv:1109.2117 [hep-ph]} \BibitemShut {NoStop}%
%%CITATION = ARXIV:1109.2117;%%
\bibitem [{\citenamefont {Watson}\ and\ \citenamefont
  {Reinhardt}(2012)}]{Watson:2011kv}%
  \BibitemOpen
  \bibfield  {author} {\bibinfo {author} {\bibfnamefont {P.}~\bibnamefont
  {Watson}}\ and\ \bibinfo {author} {\bibfnamefont {H.}~\bibnamefont
  {Reinhardt}},\ }\href {\doibase 10.1103/PhysRevD.85.025014} {\bibfield
  {journal} {\bibinfo  {journal} {Phys.~Rev.~}\ }\textbf {\bibinfo {volume}
  {D85}},\ \bibinfo {pages} {025014} (\bibinfo {year} {2012})},\ \Eprint
  {http://arxiv.org/abs/1111.6078} {arXiv:1111.6078 [hep-ph]} \BibitemShut
  {NoStop}%
%%CITATION = ARXIV:1111.6078;%%
\bibitem [{\citenamefont {Kondo}(2011)}]{Kondo:2011ab}%
  \BibitemOpen
  \bibfield  {author} {\bibinfo {author} {\bibfnamefont {K.-I.}\ \bibnamefont
  {Kondo}},\ }\href {\doibase 10.1103/PhysRevD.84.061702} {\bibfield  {journal}
  {\bibinfo  {journal} {Phys.~Rev.~}\ }\textbf {\bibinfo {volume} {D84}},\
  \bibinfo {pages} {061702} (\bibinfo {year} {2011})},\ \Eprint
  {http://arxiv.org/abs/1103.3829} {arXiv:1103.3829 [hep-th]} \BibitemShut
  {NoStop}%
%%CITATION = ARXIV:1103.3829;%%
\bibitem [{\citenamefont {Serreau}\ and\ \citenamefont
  {Tissier}(2012)}]{Serreau:2012cg}%
  \BibitemOpen
  \bibfield  {author} {\bibinfo {author} {\bibfnamefont {J.}~\bibnamefont
  {Serreau}}\ and\ \bibinfo {author} {\bibfnamefont {M.}~\bibnamefont
  {Tissier}},\ }\href {\doibase 10.1016/j.physletb.2012.04.041} {\bibfield
  {journal} {\bibinfo  {journal} {Phys. Lett.}\ }\textbf {\bibinfo {volume}
  {B712}},\ \bibinfo {pages} {97} (\bibinfo {year} {2012})},\ \Eprint
  {http://arxiv.org/abs/1202.3432} {arXiv:1202.3432 [hep-th]} \BibitemShut
  {NoStop}%
%%CITATION = ARXIV:1202.3432;%%
\bibitem [{\citenamefont {Strauss}\ \emph {et~al.}(2012)\citenamefont
  {Strauss}, \citenamefont {Fischer},\ and\ \citenamefont
  {Kellermann}}]{Strauss:2012dg}%
  \BibitemOpen
  \bibfield  {author} {\bibinfo {author} {\bibfnamefont {S.}~\bibnamefont
  {Strauss}}, \bibinfo {author} {\bibfnamefont {C.~S.}\ \bibnamefont
  {Fischer}}, \ and\ \bibinfo {author} {\bibfnamefont {C.}~\bibnamefont
  {Kellermann}},\ }\href {\doibase 10.1103/PhysRevLett.109.252001} {\bibfield
  {journal} {\bibinfo  {journal} {Phys. Rev. Lett.}\ }\textbf {\bibinfo
  {volume} {109}},\ \bibinfo {pages} {252001} (\bibinfo {year} {2012})},\
  \Eprint {http://arxiv.org/abs/1208.6239} {arXiv:1208.6239 [hep-ph]}
  \BibitemShut {NoStop}%
%%CITATION = ARXIV:1208.6239;%%
\bibitem [{\citenamefont {Siringo}(2014)}]{Siringo:2014lva}%
  \BibitemOpen
  \bibfield  {author} {\bibinfo {author} {\bibfnamefont {F.}~\bibnamefont
  {Siringo}},\ }\href {\doibase 10.1103/PhysRevD.90.094021} {\bibfield
  {journal} {\bibinfo  {journal} {Phys. Rev.}\ }\textbf {\bibinfo {volume}
  {D90}},\ \bibinfo {pages} {094021} (\bibinfo {year} {2014})},\ \Eprint
  {http://arxiv.org/abs/1408.5313} {arXiv:1408.5313 [hep-ph]} \BibitemShut
  {NoStop}%
%%CITATION = ARXIV:1408.5313;%%
\bibitem [{\citenamefont {Binosi}\ \emph {et~al.}(2015)\citenamefont {Binosi},
  \citenamefont {Chang}, \citenamefont {Papavassiliou},\ and\ \citenamefont
  {Roberts}}]{Binosi:2014aea}%
  \BibitemOpen
  \bibfield  {author} {\bibinfo {author} {\bibfnamefont {D.}~\bibnamefont
  {Binosi}}, \bibinfo {author} {\bibfnamefont {L.}~\bibnamefont {Chang}},
  \bibinfo {author} {\bibfnamefont {J.}~\bibnamefont {Papavassiliou}}, \ and\
  \bibinfo {author} {\bibfnamefont {C.~D.}\ \bibnamefont {Roberts}},\ }\href
  {\doibase 10.1016/j.physletb.2015.01.031} {\bibfield  {journal} {\bibinfo
  {journal} {Phys. Lett.}\ }\textbf {\bibinfo {volume} {B742}},\ \bibinfo
  {pages} {183} (\bibinfo {year} {2015})},\ \Eprint
  {http://arxiv.org/abs/1412.4782} {arXiv:1412.4782 [nucl-th]} \BibitemShut
  {NoStop}%
%%CITATION = ARXIV:1412.4782;%%
\bibitem [{\citenamefont {Aguilar}\ \emph {et~al.}(2015)\citenamefont
  {Aguilar}, \citenamefont {Binosi},\ and\ \citenamefont
  {Papavassiliou}}]{Aguilar:2015nqa}%
  \BibitemOpen
  \bibfield  {author} {\bibinfo {author} {\bibfnamefont {A.}~\bibnamefont
  {Aguilar}}, \bibinfo {author} {\bibfnamefont {D.}~\bibnamefont {Binosi}}, \
  and\ \bibinfo {author} {\bibfnamefont {J.}~\bibnamefont {Papavassiliou}},\
  }\href {\doibase 10.1103/PhysRevD.91.085014} {\bibfield  {journal} {\bibinfo
  {journal} {Phys.~Rev.~}\ }\textbf {\bibinfo {volume} {D91}},\ \bibinfo
  {pages} {085014} (\bibinfo {year} {2015})},\ \Eprint
  {http://arxiv.org/abs/1501.07150} {arXiv:1501.07150 [hep-ph]} \BibitemShut
  {NoStop}%
%%CITATION = ARXIV:1501.07150;%%
\bibitem [{\citenamefont {Huber}(2015)}]{Huber:2015ria}%
  \BibitemOpen
  \bibfield  {author} {\bibinfo {author} {\bibfnamefont {M.~Q.}\ \bibnamefont
  {Huber}},\ }\href {\doibase 10.1103/PhysRevD.91.085018} {\bibfield  {journal}
  {\bibinfo  {journal} {Phys.~Rev.~}\ }\textbf {\bibinfo {volume} {D91}},\
  \bibinfo {pages} {085018} (\bibinfo {year} {2015})},\ \Eprint
  {http://arxiv.org/abs/1502.04057} {arXiv:1502.04057 [hep-ph]} \BibitemShut
  {NoStop}%
%%CITATION = ARXIV:1502.04057;%%
\bibitem [{\citenamefont {Capri}\ \emph {et~al.}(2015)\citenamefont {Capri},
  \citenamefont {Dudal}, \citenamefont {Fiorentini}, \citenamefont {Guimaraes},
  \citenamefont {Justo}, \citenamefont {Pereira}, \citenamefont {Mintz},
  \citenamefont {Palhares}, \citenamefont {Sobreiro},\ and\ \citenamefont
  {Sorella}}]{Capri:2015ixa}%
  \BibitemOpen
  \bibfield  {author} {\bibinfo {author} {\bibfnamefont {M.~A.~L.}\
  \bibnamefont {Capri}}, \bibinfo {author} {\bibfnamefont {D.}~\bibnamefont
  {Dudal}}, \bibinfo {author} {\bibfnamefont {D.}~\bibnamefont {Fiorentini}},
  \bibinfo {author} {\bibfnamefont {M.~S.}\ \bibnamefont {Guimaraes}}, \bibinfo
  {author} {\bibfnamefont {I.~F.}\ \bibnamefont {Justo}}, \bibinfo {author}
  {\bibfnamefont {A.~D.}\ \bibnamefont {Pereira}}, \bibinfo {author}
  {\bibfnamefont {B.~W.}\ \bibnamefont {Mintz}}, \bibinfo {author}
  {\bibfnamefont {L.~F.}\ \bibnamefont {Palhares}}, \bibinfo {author}
  {\bibfnamefont {R.~F.}\ \bibnamefont {Sobreiro}}, \ and\ \bibinfo {author}
  {\bibfnamefont {S.~P.}\ \bibnamefont {Sorella}},\ }\href {\doibase
  10.1103/PhysRevD.92.045039} {\bibfield  {journal} {\bibinfo  {journal} {Phys.
  Rev.}\ }\textbf {\bibinfo {volume} {D92}},\ \bibinfo {pages} {045039}
  (\bibinfo {year} {2015})},\ \Eprint {http://arxiv.org/abs/1506.06995}
  {arXiv:1506.06995 [hep-th]} \BibitemShut {NoStop}%
%%CITATION = ARXIV:1506.06995;%%
\bibitem [{\citenamefont {Binosi}\ \emph {et~al.}(2017)\citenamefont {Binosi},
  \citenamefont {Mezrag}, \citenamefont {Papavassiliou}, \citenamefont
  {Roberts},\ and\ \citenamefont {Rodriguez-Quintero}}]{Binosi:2016nme}%
  \BibitemOpen
  \bibfield  {author} {\bibinfo {author} {\bibfnamefont {D.}~\bibnamefont
  {Binosi}}, \bibinfo {author} {\bibfnamefont {C.}~\bibnamefont {Mezrag}},
  \bibinfo {author} {\bibfnamefont {J.}~\bibnamefont {Papavassiliou}}, \bibinfo
  {author} {\bibfnamefont {C.~D.}\ \bibnamefont {Roberts}}, \ and\ \bibinfo
  {author} {\bibfnamefont {J.}~\bibnamefont {Rodriguez-Quintero}},\ }\href
  {\doibase 10.1103/PhysRevD.96.054026} {\bibfield  {journal} {\bibinfo
  {journal} {Phys. Rev.}\ }\textbf {\bibinfo {volume} {D96}},\ \bibinfo {pages}
  {054026} (\bibinfo {year} {2017})},\ \Eprint
  {http://arxiv.org/abs/1612.04835} {arXiv:1612.04835 [nucl-th]} \BibitemShut
  {NoStop}%
%%CITATION = ARXIV:1612.04835;%%
\bibitem [{\citenamefont {Glazek}\ \emph {et~al.}(2017)\citenamefont
  {Głazek}, \citenamefont {G\'omez-Rocha}, \citenamefont {More},\ and\
  \citenamefont {Serafin}}]{Glazek:2017rwe}%
  \BibitemOpen
  \bibfield  {author} {\bibinfo {author} {\bibfnamefont {S.~D.}\ \bibnamefont
  {Glazek}}, \bibinfo {author} {\bibfnamefont {M.}~\bibnamefont
  {G\'omez-Rocha}}, \bibinfo {author} {\bibfnamefont {J.}~\bibnamefont {More}},
  \ and\ \bibinfo {author} {\bibfnamefont {K.}~\bibnamefont {Serafin}},\ }\href
  {\doibase 10.1016/j.physletb.2017.08.018} {\bibfield  {journal} {\bibinfo
  {journal} {Phys. Lett.}\ }\textbf {\bibinfo {volume} {B773}},\ \bibinfo
  {pages} {172} (\bibinfo {year} {2017})},\ \Eprint
  {http://arxiv.org/abs/1705.07629} {arXiv:1705.07629 [hep-ph]} \BibitemShut
  {NoStop}%
%%CITATION = ARXIV:1705.07629;%%
\bibitem [{\citenamefont {Gao}\ \emph {et~al.}(2017)\citenamefont {Gao},
  \citenamefont {Qin}, \citenamefont {Roberts},\ and\ \citenamefont
  {Rodriguez-Quintero}}]{Gao:2017uox}%
  \BibitemOpen
  \bibfield  {author} {\bibinfo {author} {\bibfnamefont {F.}~\bibnamefont
  {Gao}}, \bibinfo {author} {\bibfnamefont {S.-X.}\ \bibnamefont {Qin}},
  \bibinfo {author} {\bibfnamefont {C.~D.}\ \bibnamefont {Roberts}}, \ and\
  \bibinfo {author} {\bibfnamefont {J.}~\bibnamefont {Rodriguez-Quintero}},\
  }\href@noop {} {\  (\bibinfo {year} {2017})},\ \Eprint
  {http://arxiv.org/abs/1706.04681} {arXiv:1706.04681 [hep-ph]} \BibitemShut
  {NoStop}%
%%CITATION = ARXIV:1706.04681;%%
\bibitem [{\citenamefont {Cornwall}\ and\ \citenamefont
  {Papavassiliou}(1989)}]{Cornwall:1989gv}%
  \BibitemOpen
  \bibfield  {author} {\bibinfo {author} {\bibfnamefont {J.~M.}\ \bibnamefont
  {Cornwall}}\ and\ \bibinfo {author} {\bibfnamefont {J.}~\bibnamefont
  {Papavassiliou}},\ }\href@noop {} {\bibfield  {journal} {\bibinfo  {journal}
  {Phys. Rev.}\ }\textbf {\bibinfo {volume} {D40}},\ \bibinfo {pages} {3474}
  (\bibinfo {year} {1989})}\BibitemShut {NoStop}%
%%CITATION = PHRVA,D40,3474;%%
\bibitem [{\citenamefont {Pilaftsis}(1997)}]{Pilaftsis:1996fh}%
  \BibitemOpen
  \bibfield  {author} {\bibinfo {author} {\bibfnamefont {A.}~\bibnamefont
  {Pilaftsis}},\ }\href@noop {} {\bibfield  {journal} {\bibinfo  {journal}
  {Nucl. Phys.}\ }\textbf {\bibinfo {volume} {B487}},\ \bibinfo {pages} {467}
  (\bibinfo {year} {1997})},\ \Eprint {http://arxiv.org/abs/hep-ph/9607451}
  {hep-ph/9607451} \BibitemShut {NoStop}%
%%CITATION = HEP-PH/9607451;%%
\bibitem [{\citenamefont {Binosi}\ and\ \citenamefont
  {Papavassiliou}(2002{\natexlab{a}})}]{Binosi:2002ft}%
  \BibitemOpen
  \bibfield  {author} {\bibinfo {author} {\bibfnamefont {D.}~\bibnamefont
  {Binosi}}\ and\ \bibinfo {author} {\bibfnamefont {J.}~\bibnamefont
  {Papavassiliou}},\ }\href@noop {} {\bibfield  {journal} {\bibinfo  {journal}
  {Phys. Rev.}\ }\textbf {\bibinfo {volume} {D66}},\ \bibinfo {pages}
  {111901(R)} (\bibinfo {year} {2002}{\natexlab{a}})},\ \Eprint
  {http://arxiv.org/abs/hep-ph/0208189} {hep-ph/0208189} \BibitemShut {NoStop}%
%%CITATION = HEP-PH/0208189;%%
\bibitem [{\citenamefont {Binosi}\ and\ \citenamefont
  {Papavassiliou}(2004)}]{Binosi:2003rr}%
  \BibitemOpen
  \bibfield  {author} {\bibinfo {author} {\bibfnamefont {D.}~\bibnamefont
  {Binosi}}\ and\ \bibinfo {author} {\bibfnamefont {J.}~\bibnamefont
  {Papavassiliou}},\ }\href {\doibase 10.1088/0954-3899/30/2/017} {\bibfield
  {journal} {\bibinfo  {journal} {J.Phys.G}\ }\textbf {\bibinfo {volume}
  {G30}},\ \bibinfo {pages} {203} (\bibinfo {year} {2004})},\ \Eprint
  {http://arxiv.org/abs/hep-ph/0301096} {arXiv:hep-ph/0301096 [hep-ph]}
  \BibitemShut {NoStop}%
\bibitem [{\citenamefont {Binosi}\ and\ \citenamefont
  {Papavassiliou}(2009)}]{Binosi:2009qm}%
  \BibitemOpen
  \bibfield  {author} {\bibinfo {author} {\bibfnamefont {D.}~\bibnamefont
  {Binosi}}\ and\ \bibinfo {author} {\bibfnamefont {J.}~\bibnamefont
  {Papavassiliou}},\ }\href {\doibase 10.1016/j.physrep.2009.05.001} {\bibfield
   {journal} {\bibinfo  {journal} {Phys. Rept.}\ }\textbf {\bibinfo {volume}
  {479}},\ \bibinfo {pages} {1} (\bibinfo {year} {2009})},\ \Eprint
  {http://arxiv.org/abs/0909.2536} {arXiv:0909.2536 [hep-ph]} \BibitemShut
  {NoStop}%
%%CITATION = ARXIV:0909.2536;%%
\bibitem [{\citenamefont {Abbott}(1981)}]{Abbott:1980hw}%
  \BibitemOpen
  \bibfield  {author} {\bibinfo {author} {\bibfnamefont {L.~F.}\ \bibnamefont
  {Abbott}},\ }\href@noop {} {\bibfield  {journal} {\bibinfo  {journal} {Nucl.
  Phys.}\ }\textbf {\bibinfo {volume} {B185}},\ \bibinfo {pages} {189}
  (\bibinfo {year} {1981})}\BibitemShut {NoStop}%
%%CITATION = NUPHA,B185,189;%%
\bibitem [{\citenamefont {Binosi}\ and\ \citenamefont
  {Papavassiliou}(2008{\natexlab{a}})}]{Binosi:2007pi}%
  \BibitemOpen
  \bibfield  {author} {\bibinfo {author} {\bibfnamefont {D.}~\bibnamefont
  {Binosi}}\ and\ \bibinfo {author} {\bibfnamefont {J.}~\bibnamefont
  {Papavassiliou}},\ }\href {\doibase 10.1103/PhysRevD.77.061702} {\bibfield
  {journal} {\bibinfo  {journal} {Phys.~Rev.~}\ }\textbf {\bibinfo {volume}
  {D77}},\ \bibinfo {pages} {061702} (\bibinfo {year} {2008}{\natexlab{a}})},\
  \Eprint {http://arxiv.org/abs/0712.2707} {arXiv:0712.2707 [hep-ph]}
  \BibitemShut {NoStop}%
\bibitem [{\citenamefont {Binosi}\ and\ \citenamefont
  {Papavassiliou}(2008{\natexlab{b}})}]{Binosi:2008qk}%
  \BibitemOpen
  \bibfield  {author} {\bibinfo {author} {\bibfnamefont {D.}~\bibnamefont
  {Binosi}}\ and\ \bibinfo {author} {\bibfnamefont {J.}~\bibnamefont
  {Papavassiliou}},\ }\href {\doibase 10.1088/1126-6708/2008/11/063} {\bibfield
   {journal} {\bibinfo  {journal} {JHEP}\ }\textbf {\bibinfo {volume} {0811}},\
  \bibinfo {pages} {063} (\bibinfo {year} {2008}{\natexlab{b}})},\ \Eprint
  {http://arxiv.org/abs/0805.3994} {arXiv:0805.3994 [hep-ph]} \BibitemShut
  {NoStop}%
\bibitem [{\citenamefont {Aguilar}\ and\ \citenamefont
  {Papavassiliou}(2010)}]{Aguilar:2009ke}%
  \BibitemOpen
  \bibfield  {author} {\bibinfo {author} {\bibfnamefont {A.~C.}\ \bibnamefont
  {Aguilar}}\ and\ \bibinfo {author} {\bibfnamefont {J.}~\bibnamefont
  {Papavassiliou}},\ }\href {\doibase 10.1103/PhysRevD.81.034003} {\bibfield
  {journal} {\bibinfo  {journal} {Phys.~Rev.~}\ }\textbf {\bibinfo {volume}
  {D81}},\ \bibinfo {pages} {034003} (\bibinfo {year} {2010})},\ \Eprint
  {http://arxiv.org/abs/0910.4142} {arXiv:0910.4142 [hep-ph]} \BibitemShut
  {NoStop}%
\bibitem [{\citenamefont {Schwinger}(1962{\natexlab{a}})}]{Schwinger:1962tn}%
  \BibitemOpen
  \bibfield  {author} {\bibinfo {author} {\bibfnamefont {J.~S.}\ \bibnamefont
  {Schwinger}},\ }\href@noop {} {\bibfield  {journal} {\bibinfo  {journal}
  {Phys. Rev.}\ }\textbf {\bibinfo {volume} {125}},\ \bibinfo {pages} {397}
  (\bibinfo {year} {1962}{\natexlab{a}})}\BibitemShut {NoStop}%
%%CITATION = PHRVA,125,397;%%
\bibitem [{\citenamefont {Schwinger}(1962{\natexlab{b}})}]{Schwinger:1962tp}%
  \BibitemOpen
  \bibfield  {author} {\bibinfo {author} {\bibfnamefont {J.~S.}\ \bibnamefont
  {Schwinger}},\ }\href@noop {} {\bibfield  {journal} {\bibinfo  {journal}
  {Phys. Rev.}\ }\textbf {\bibinfo {volume} {128}},\ \bibinfo {pages} {2425}
  (\bibinfo {year} {1962}{\natexlab{b}})}\BibitemShut {NoStop}%
%%CITATION = PHRVA,128,2425;%%
\bibitem [{\citenamefont {Jackiw}\ and\ \citenamefont
  {Johnson}(1973)}]{Jackiw:1973tr}%
  \BibitemOpen
  \bibfield  {author} {\bibinfo {author} {\bibfnamefont {R.}~\bibnamefont
  {Jackiw}}\ and\ \bibinfo {author} {\bibfnamefont {K.}~\bibnamefont
  {Johnson}},\ }\href@noop {} {\bibfield  {journal} {\bibinfo  {journal} {Phys.
  Rev.}\ }\textbf {\bibinfo {volume} {D8}},\ \bibinfo {pages} {2386} (\bibinfo
  {year} {1973})}\BibitemShut {NoStop}%
%%CITATION = PHRVA,D8,2386;%%
\bibitem [{\citenamefont {Smit}(1974)}]{Smit:1974je}%
  \BibitemOpen
  \bibfield  {author} {\bibinfo {author} {\bibfnamefont {J.}~\bibnamefont
  {Smit}},\ }\href {\doibase 10.1103/PhysRevD.10.2473} {\bibfield  {journal}
  {\bibinfo  {journal} {Phys. Rev.}\ }\textbf {\bibinfo {volume} {D10}},\
  \bibinfo {pages} {2473} (\bibinfo {year} {1974})}\BibitemShut {NoStop}%
%%CITATION = PHRVA,D10,2473;%%
\bibitem [{\citenamefont {Eichten}\ and\ \citenamefont
  {Feinberg}(1974)}]{Eichten:1974et}%
  \BibitemOpen
  \bibfield  {author} {\bibinfo {author} {\bibfnamefont {E.}~\bibnamefont
  {Eichten}}\ and\ \bibinfo {author} {\bibfnamefont {F.}~\bibnamefont
  {Feinberg}},\ }\href@noop {} {\bibfield  {journal} {\bibinfo  {journal}
  {Phys. Rev.}\ }\textbf {\bibinfo {volume} {D10}},\ \bibinfo {pages} {3254}
  (\bibinfo {year} {1974})}\BibitemShut {NoStop}%
%%CITATION = PHRVA,D10,3254;%%
\bibitem [{\citenamefont {Poggio}\ \emph {et~al.}(1975)\citenamefont {Poggio},
  \citenamefont {Tomboulis},\ and\ \citenamefont {Tye}}]{Poggio:1974qs}%
  \BibitemOpen
  \bibfield  {author} {\bibinfo {author} {\bibfnamefont {E.~C.}\ \bibnamefont
  {Poggio}}, \bibinfo {author} {\bibfnamefont {E.}~\bibnamefont {Tomboulis}}, \
  and\ \bibinfo {author} {\bibfnamefont {S.~H.~H.}\ \bibnamefont {Tye}},\
  }\href {\doibase 10.1103/PhysRevD.11.2839} {\bibfield  {journal} {\bibinfo
  {journal} {Phys. Rev.}\ }\textbf {\bibinfo {volume} {D11}},\ \bibinfo {pages}
  {2839} (\bibinfo {year} {1975})}\BibitemShut {NoStop}%
%%CITATION = PHRVA,D11,2839;%%
\bibitem [{\citenamefont {Binosi}\ and\ \citenamefont
  {Papavassiliou}(2017)}]{Binosi:2017rwj}%
  \BibitemOpen
  \bibfield  {author} {\bibinfo {author} {\bibfnamefont {D.}~\bibnamefont
  {Binosi}}\ and\ \bibinfo {author} {\bibfnamefont {J.}~\bibnamefont
  {Papavassiliou}},\ }\href@noop {} {\  (\bibinfo {year} {2017})},\ \Eprint
  {http://arxiv.org/abs/1709.09964} {arXiv:1709.09964 [hep-ph]} \BibitemShut
  {NoStop}%
%%CITATION = ARXIV:1709.09964;%%
\bibitem [{\citenamefont {Taylor}(1971)}]{Taylor:1971ff}%
  \BibitemOpen
  \bibfield  {author} {\bibinfo {author} {\bibfnamefont {J.~C.}\ \bibnamefont
  {Taylor}},\ }\href@noop {} {\bibfield  {journal} {\bibinfo  {journal} {Nucl.
  Phys.}\ }\textbf {\bibinfo {volume} {B33}},\ \bibinfo {pages} {436} (\bibinfo
  {year} {1971})}\BibitemShut {NoStop}%
%%CITATION = NUPHA,B33,436;%%
\bibitem [{\citenamefont {Binosi}\ and\ \citenamefont
  {Papavassiliou}(2002{\natexlab{b}})}]{Binosi:2002ez}%
  \BibitemOpen
  \bibfield  {author} {\bibinfo {author} {\bibfnamefont {D.}~\bibnamefont
  {Binosi}}\ and\ \bibinfo {author} {\bibfnamefont {J.}~\bibnamefont
  {Papavassiliou}},\ }\href {\doibase 10.1103/PhysRevD.66.025024} {\bibfield
  {journal} {\bibinfo  {journal} {Phys.~Rev.~}\ }\textbf {\bibinfo {volume}
  {D66}},\ \bibinfo {pages} {025024} (\bibinfo {year} {2002}{\natexlab{b}})},\
  \Eprint {http://arxiv.org/abs/hep-ph/0204128} {arXiv:hep-ph/0204128 [hep-ph]}
  \BibitemShut {NoStop}%
\bibitem [{\citenamefont {Binosi}\ and\ \citenamefont
  {Quadri}(2013)}]{Binosi:2013cea}%
  \BibitemOpen
  \bibfield  {author} {\bibinfo {author} {\bibfnamefont {D.}~\bibnamefont
  {Binosi}}\ and\ \bibinfo {author} {\bibfnamefont {A.}~\bibnamefont
  {Quadri}},\ }\href {\doibase 10.1103/PhysRevD.88.085036} {\bibfield
  {journal} {\bibinfo  {journal} {Phys.~Rev.~}\ }\textbf {\bibinfo {volume}
  {D88}},\ \bibinfo {pages} {085036} (\bibinfo {year} {2013})},\ \Eprint
  {http://arxiv.org/abs/1309.1021} {arXiv:1309.1021 [hep-th]} \BibitemShut
  {NoStop}%
%%CITATION = ARXIV:1309.1021;%%
\bibitem [{\citenamefont {Grassi}\ \emph {et~al.}(2004)\citenamefont {Grassi},
  \citenamefont {Hurth},\ and\ \citenamefont {Quadri}}]{Grassi:2004yq}%
  \BibitemOpen
  \bibfield  {author} {\bibinfo {author} {\bibfnamefont {P.~A.}\ \bibnamefont
  {Grassi}}, \bibinfo {author} {\bibfnamefont {T.}~\bibnamefont {Hurth}}, \
  and\ \bibinfo {author} {\bibfnamefont {A.}~\bibnamefont {Quadri}},\
  }\href@noop {} {\bibfield  {journal} {\bibinfo  {journal} {Phys. Rev.}\
  }\textbf {\bibinfo {volume} {D70}},\ \bibinfo {pages} {105014} (\bibinfo
  {year} {2004})},\ \Eprint {http://arxiv.org/abs/hep-th/0405104}
  {hep-th/0405104} \BibitemShut {NoStop}%
%%CITATION = HEP-TH/0405104;%%
\bibitem [{\citenamefont {Binosi}\ \emph {et~al.}(2012)\citenamefont {Binosi},
  \citenamefont {Iba\~nez},\ and\ \citenamefont
  {Papavassiliou}}]{Binosi:2012sj}%
  \BibitemOpen
  \bibfield  {author} {\bibinfo {author} {\bibfnamefont {D.}~\bibnamefont
  {Binosi}}, \bibinfo {author} {\bibfnamefont {D.}~\bibnamefont {Iba\~nez}}, \
  and\ \bibinfo {author} {\bibfnamefont {J.}~\bibnamefont {Papavassiliou}},\
  }\href {\doibase 10.1103/PhysRevD.86.085033} {\bibfield  {journal} {\bibinfo
  {journal} {Phys. Rev.}\ }\textbf {\bibinfo {volume} {D86}},\ \bibinfo {pages}
  {085033} (\bibinfo {year} {2012})},\ \Eprint {http://arxiv.org/abs/1208.1451}
  {arXiv:1208.1451 [hep-ph]} \BibitemShut {NoStop}%
%%CITATION = ARXIV:1208.1451;%%
\bibitem [{\citenamefont {Aguilar}\ \emph {et~al.}(2013)\citenamefont
  {Aguilar}, \citenamefont {Iba{\~n}ez},\ and\ \citenamefont
  {Papavassiliou}}]{Aguilar:2013xqa}%
  \BibitemOpen
  \bibfield  {author} {\bibinfo {author} {\bibfnamefont {A.~C.}\ \bibnamefont
  {Aguilar}}, \bibinfo {author} {\bibfnamefont {D.}~\bibnamefont {Iba{\~n}ez}},
  \ and\ \bibinfo {author} {\bibfnamefont {J.}~\bibnamefont {Papavassiliou}},\
  }\href {\doibase 10.1103/PhysRevD.87.114020} {\bibfield  {journal} {\bibinfo
  {journal} {Phys. Rev.}\ }\textbf {\bibinfo {volume} {D87}},\ \bibinfo {pages}
  {114020} (\bibinfo {year} {2013})},\ \Eprint {http://arxiv.org/abs/1303.3609}
  {arXiv:1303.3609 [hep-ph]} \BibitemShut {NoStop}%
%%CITATION = ARXIV:1303.3609;%%
\bibitem [{\citenamefont {Ball}\ and\ \citenamefont
  {Chiu}(1980)}]{Ball:1980ay}%
  \BibitemOpen
  \bibfield  {author} {\bibinfo {author} {\bibfnamefont {J.~S.}\ \bibnamefont
  {Ball}}\ and\ \bibinfo {author} {\bibfnamefont {T.-W.}\ \bibnamefont
  {Chiu}},\ }\href {\doibase 10.1103/PhysRevD.22.2542} {\bibfield  {journal}
  {\bibinfo  {journal} {Phys.~Rev.~}\ }\textbf {\bibinfo {volume} {D22}},\
  \bibinfo {pages} {2542} (\bibinfo {year} {1980})}\BibitemShut {NoStop}%
%%CITATION = PHRVA,D22,2542;%%
\bibitem [{\citenamefont {Aguilar}\ \emph {et~al.}(2017)\citenamefont
  {Aguilar}, \citenamefont {Binosi},\ and\ \citenamefont
  {Papavassiliou}}]{Aguilar:2016ock}%
  \BibitemOpen
  \bibfield  {author} {\bibinfo {author} {\bibfnamefont {A.~C.}\ \bibnamefont
  {Aguilar}}, \bibinfo {author} {\bibfnamefont {D.}~\bibnamefont {Binosi}}, \
  and\ \bibinfo {author} {\bibfnamefont {J.}~\bibnamefont {Papavassiliou}},\
  }\href {\doibase 10.1103/PhysRevD.95.034017} {\bibfield  {journal} {\bibinfo
  {journal} {Phys. Rev.}\ }\textbf {\bibinfo {volume} {D95}},\ \bibinfo {pages}
  {034017} (\bibinfo {year} {2017})},\ \Eprint
  {http://arxiv.org/abs/1611.02096} {arXiv:1611.02096 [hep-ph]} \BibitemShut
  {NoStop}%
%%CITATION = ARXIV:1611.02096;%%
\bibitem [{\citenamefont {Athenodorou}\ \emph {et~al.}(2016)\citenamefont
  {Athenodorou}, \citenamefont {Binosi}, \citenamefont {Boucaud}, \citenamefont
  {De~Soto}, \citenamefont {Papavassiliou}, \citenamefont
  {Rodriguez-Quintero},\ and\ \citenamefont
  {Zafeiropoulos}}]{Athenodorou:2016oyh}%
  \BibitemOpen
  \bibfield  {author} {\bibinfo {author} {\bibfnamefont {A.}~\bibnamefont
  {Athenodorou}}, \bibinfo {author} {\bibfnamefont {D.}~\bibnamefont {Binosi}},
  \bibinfo {author} {\bibfnamefont {P.}~\bibnamefont {Boucaud}}, \bibinfo
  {author} {\bibfnamefont {F.}~\bibnamefont {De~Soto}}, \bibinfo {author}
  {\bibfnamefont {J.}~\bibnamefont {Papavassiliou}}, \bibinfo {author}
  {\bibfnamefont {J.}~\bibnamefont {Rodriguez-Quintero}}, \ and\ \bibinfo
  {author} {\bibfnamefont {S.}~\bibnamefont {Zafeiropoulos}},\ }\href {\doibase
  10.1016/j.physletb.2016.08.065} {\bibfield  {journal} {\bibinfo  {journal}
  {Phys. Lett.}\ }\textbf {\bibinfo {volume} {B761}},\ \bibinfo {pages} {444}
  (\bibinfo {year} {2016})},\ \Eprint {http://arxiv.org/abs/1607.01278}
  {arXiv:1607.01278 [hep-ph]} \BibitemShut {NoStop}%
%%CITATION = ARXIV:1607.01278;%%
\bibitem [{\citenamefont {Boucaud}\ \emph {et~al.}(2017)\citenamefont
  {Boucaud}, \citenamefont {De~Soto}, \citenamefont {Rodríguez-Quintero},\
  and\ \citenamefont {Zafeiropoulos}}]{Boucaud:2017obn}%
  \BibitemOpen
  \bibfield  {author} {\bibinfo {author} {\bibfnamefont {P.}~\bibnamefont
  {Boucaud}}, \bibinfo {author} {\bibfnamefont {F.}~\bibnamefont {De~Soto}},
  \bibinfo {author} {\bibfnamefont {J.}~\bibnamefont {Rodríguez-Quintero}}, \
  and\ \bibinfo {author} {\bibfnamefont {S.}~\bibnamefont {Zafeiropoulos}},\
  }\href {\doibase 10.1103/PhysRevD.95.114503} {\bibfield  {journal} {\bibinfo
  {journal} {Phys. Rev.}\ }\textbf {\bibinfo {volume} {D95}},\ \bibinfo {pages}
  {114503} (\bibinfo {year} {2017})},\ \Eprint
  {http://arxiv.org/abs/1701.07390} {arXiv:1701.07390 [hep-lat]} \BibitemShut
  {NoStop}%
%%CITATION = ARXIV:1701.07390;%%
\bibitem [{\citenamefont {Alkofer}\ \emph {et~al.}(2009)\citenamefont
  {Alkofer}, \citenamefont {Huber},\ and\ \citenamefont
  {Schwenzer}}]{Alkofer:2008dt}%
  \BibitemOpen
  \bibfield  {author} {\bibinfo {author} {\bibfnamefont {R.}~\bibnamefont
  {Alkofer}}, \bibinfo {author} {\bibfnamefont {M.~Q.}\ \bibnamefont {Huber}},
  \ and\ \bibinfo {author} {\bibfnamefont {K.}~\bibnamefont {Schwenzer}},\
  }\href {\doibase 10.1140/epjc/s10052-009-1066-3} {\bibfield  {journal}
  {\bibinfo  {journal} {Eur. Phys. J.}\ }\textbf {\bibinfo {volume} {C62}},\
  \bibinfo {pages} {761} (\bibinfo {year} {2009})},\ \Eprint
  {http://arxiv.org/abs/0812.4045} {arXiv:0812.4045 [hep-ph]} \BibitemShut
  {NoStop}%
%%CITATION = ARXIV:0812.4045;%%
\bibitem [{\citenamefont {Tissier}\ and\ \citenamefont
  {Wschebor}(2011)}]{Tissier:2011ey}%
  \BibitemOpen
  \bibfield  {author} {\bibinfo {author} {\bibfnamefont {M.}~\bibnamefont
  {Tissier}}\ and\ \bibinfo {author} {\bibfnamefont {N.}~\bibnamefont
  {Wschebor}},\ }\href {\doibase 10.1103/PhysRevD.84.045018} {\bibfield
  {journal} {\bibinfo  {journal} {Phys. Rev.}\ }\textbf {\bibinfo {volume}
  {D84}},\ \bibinfo {pages} {045018} (\bibinfo {year} {2011})},\ \Eprint
  {http://arxiv.org/abs/1105.2475} {arXiv:1105.2475 [hep-th]} \BibitemShut
  {NoStop}%
%%CITATION = ARXIV:1105.2475;%%
\bibitem [{\citenamefont {Pelaez}\ \emph {et~al.}(2013)\citenamefont {Pelaez},
  \citenamefont {Tissier},\ and\ \citenamefont {Wschebor}}]{Pelaez:2013cpa}%
  \BibitemOpen
  \bibfield  {author} {\bibinfo {author} {\bibfnamefont {M.}~\bibnamefont
  {Pelaez}}, \bibinfo {author} {\bibfnamefont {M.}~\bibnamefont {Tissier}}, \
  and\ \bibinfo {author} {\bibfnamefont {N.}~\bibnamefont {Wschebor}},\ }\href
  {\doibase 10.1103/PhysRevD.88.125003} {\bibfield  {journal} {\bibinfo
  {journal} {Phys.~Rev.~}\ }\textbf {\bibinfo {volume} {D88}},\ \bibinfo
  {pages} {125003} (\bibinfo {year} {2013})},\ \Eprint
  {http://arxiv.org/abs/1310.2594} {arXiv:1310.2594 [hep-th]} \BibitemShut
  {NoStop}%
%%CITATION = ARXIV:1310.2594;%%
\bibitem [{\citenamefont {Aguilar}\ \emph
  {et~al.}(2014{\natexlab{a}})\citenamefont {Aguilar}, \citenamefont {Binosi},
  \citenamefont {Iba{\~n}ez},\ and\ \citenamefont
  {Papavassiliou}}]{Aguilar:2013vaa}%
  \BibitemOpen
  \bibfield  {author} {\bibinfo {author} {\bibfnamefont {A.~C.}\ \bibnamefont
  {Aguilar}}, \bibinfo {author} {\bibfnamefont {D.}~\bibnamefont {Binosi}},
  \bibinfo {author} {\bibfnamefont {D.}~\bibnamefont {Iba{\~n}ez}}, \ and\
  \bibinfo {author} {\bibfnamefont {J.}~\bibnamefont {Papavassiliou}},\ }\href
  {\doibase 10.1103/PhysRevD.89.085008} {\bibfield  {journal} {\bibinfo
  {journal} {Phys. Rev.}\ }\textbf {\bibinfo {volume} {D89}},\ \bibinfo {pages}
  {085008} (\bibinfo {year} {2014}{\natexlab{a}})},\ \Eprint
  {http://arxiv.org/abs/1312.1212} {arXiv:1312.1212 [hep-ph]} \BibitemShut
  {NoStop}%
%%CITATION = ARXIV:1312.1212;%%
\bibitem [{\citenamefont {Blum}\ \emph {et~al.}(2014)\citenamefont {Blum},
  \citenamefont {Huber}, \citenamefont {Mitter},\ and\ \citenamefont {von
  Smekal}}]{Blum:2014gna}%
  \BibitemOpen
  \bibfield  {author} {\bibinfo {author} {\bibfnamefont {A.}~\bibnamefont
  {Blum}}, \bibinfo {author} {\bibfnamefont {M.~Q.}\ \bibnamefont {Huber}},
  \bibinfo {author} {\bibfnamefont {M.}~\bibnamefont {Mitter}}, \ and\ \bibinfo
  {author} {\bibfnamefont {L.}~\bibnamefont {von Smekal}},\ }\href {\doibase
  10.1103/PhysRevD.89.061703} {\bibfield  {journal} {\bibinfo  {journal}
  {Phys.~Rev.~}\ }\textbf {\bibinfo {volume} {D89}},\ \bibinfo {pages} {061703}
  (\bibinfo {year} {2014})},\ \Eprint {http://arxiv.org/abs/1401.0713}
  {arXiv:1401.0713 [hep-ph]} \BibitemShut {NoStop}%
%%CITATION = ARXIV:1401.0713;%%
\bibitem [{\citenamefont {Eichmann}\ \emph {et~al.}(2014)\citenamefont
  {Eichmann}, \citenamefont {Williams}, \citenamefont {Alkofer},\ and\
  \citenamefont {Vujinovic}}]{Eichmann:2014xya}%
  \BibitemOpen
  \bibfield  {author} {\bibinfo {author} {\bibfnamefont {G.}~\bibnamefont
  {Eichmann}}, \bibinfo {author} {\bibfnamefont {R.}~\bibnamefont {Williams}},
  \bibinfo {author} {\bibfnamefont {R.}~\bibnamefont {Alkofer}}, \ and\
  \bibinfo {author} {\bibfnamefont {M.}~\bibnamefont {Vujinovic}},\ }\href
  {\doibase 10.1103/PhysRevD.89.105014} {\bibfield  {journal} {\bibinfo
  {journal} {Phys.~Rev.~}\ }\textbf {\bibinfo {volume} {D89}},\ \bibinfo
  {pages} {105014} (\bibinfo {year} {2014})},\ \Eprint
  {http://arxiv.org/abs/1402.1365} {arXiv:1402.1365 [hep-ph]} \BibitemShut
  {NoStop}%
%%CITATION = ARXIV:1402.1365;%%
\bibitem [{\citenamefont {Williams}\ \emph {et~al.}(2016)\citenamefont
  {Williams}, \citenamefont {Fischer},\ and\ \citenamefont
  {Heupel}}]{Williams:2015cvx}%
  \BibitemOpen
  \bibfield  {author} {\bibinfo {author} {\bibfnamefont {R.}~\bibnamefont
  {Williams}}, \bibinfo {author} {\bibfnamefont {C.~S.}\ \bibnamefont
  {Fischer}}, \ and\ \bibinfo {author} {\bibfnamefont {W.}~\bibnamefont
  {Heupel}},\ }\href {\doibase 10.1103/PhysRevD.93.034026} {\bibfield
  {journal} {\bibinfo  {journal} {Phys. Rev.}\ }\textbf {\bibinfo {volume}
  {D93}},\ \bibinfo {pages} {034026} (\bibinfo {year} {2016})},\ \Eprint
  {http://arxiv.org/abs/1512.00455} {arXiv:1512.00455 [hep-ph]} \BibitemShut
  {NoStop}%
%%CITATION = ARXIV:1512.00455;%%
\bibitem [{\citenamefont {Cyrol}\ \emph {et~al.}(2016)\citenamefont {Cyrol},
  \citenamefont {Fister}, \citenamefont {Mitter}, \citenamefont {Pawlowski},\
  and\ \citenamefont {Strodthoff}}]{Cyrol:2016tym}%
  \BibitemOpen
  \bibfield  {author} {\bibinfo {author} {\bibfnamefont {A.~K.}\ \bibnamefont
  {Cyrol}}, \bibinfo {author} {\bibfnamefont {L.}~\bibnamefont {Fister}},
  \bibinfo {author} {\bibfnamefont {M.}~\bibnamefont {Mitter}}, \bibinfo
  {author} {\bibfnamefont {J.~M.}\ \bibnamefont {Pawlowski}}, \ and\ \bibinfo
  {author} {\bibfnamefont {N.}~\bibnamefont {Strodthoff}},\ }\href {\doibase
  10.1103/PhysRevD.94.054005} {\bibfield  {journal} {\bibinfo  {journal} {Phys.
  Rev.}\ }\textbf {\bibinfo {volume} {D94}},\ \bibinfo {pages} {054005}
  (\bibinfo {year} {2016})},\ \Eprint {http://arxiv.org/abs/1605.01856}
  {arXiv:1605.01856 [hep-ph]} \BibitemShut {NoStop}%
%%CITATION = ARXIV:1605.01856;%%
\bibitem [{\citenamefont {Aguilar}\ \emph
  {et~al.}(2014{\natexlab{b}})\citenamefont {Aguilar}, \citenamefont {Binosi},\
  and\ \citenamefont {Papavassiliou}}]{Aguilar:2014tka}%
  \BibitemOpen
  \bibfield  {author} {\bibinfo {author} {\bibfnamefont {A.~C.}\ \bibnamefont
  {Aguilar}}, \bibinfo {author} {\bibfnamefont {D.}~\bibnamefont {Binosi}}, \
  and\ \bibinfo {author} {\bibfnamefont {J.}~\bibnamefont {Papavassiliou}},\
  }\href {\doibase 10.1103/PhysRevD.89.085032} {\bibfield  {journal} {\bibinfo
  {journal} {Phys. Rev.}\ }\textbf {\bibinfo {volume} {D89}},\ \bibinfo {pages}
  {085032} (\bibinfo {year} {2014}{\natexlab{b}})},\ \Eprint
  {http://arxiv.org/abs/1401.3631} {arXiv:1401.3631 [hep-ph]} \BibitemShut
  {NoStop}%
%%CITATION = ARXIV:1401.3631;%%
\bibitem [{\citenamefont {Pascual}\ and\ \citenamefont
  {Tarrach}(1980)}]{Pascual:1980yu}%
  \BibitemOpen
  \bibfield  {author} {\bibinfo {author} {\bibfnamefont {P.}~\bibnamefont
  {Pascual}}\ and\ \bibinfo {author} {\bibfnamefont {R.}~\bibnamefont
  {Tarrach}},\ }\href@noop {} {\bibfield  {journal} {\bibinfo  {journal} {Nucl.
  Phys.}\ }\textbf {\bibinfo {volume} {B174}},\ \bibinfo {pages} {123}
  (\bibinfo {year} {1980})}\BibitemShut {NoStop}%
%%CITATION = NUPHA,B174,123;%%
\bibitem [{\citenamefont {Binosi}\ \emph {et~al.}(2014)\citenamefont {Binosi},
  \citenamefont {Iba{\~n}ez},\ and\ \citenamefont
  {Papavassiliou}}]{Binosi:2014kka}%
  \BibitemOpen
  \bibfield  {author} {\bibinfo {author} {\bibfnamefont {D.}~\bibnamefont
  {Binosi}}, \bibinfo {author} {\bibfnamefont {D.}~\bibnamefont {Iba{\~n}ez}},
  \ and\ \bibinfo {author} {\bibfnamefont {J.}~\bibnamefont {Papavassiliou}},\
  }\href {\doibase 10.1007/JHEP09(2014)059} {\bibfield  {journal} {\bibinfo
  {journal} {JHEP}\ }\textbf {\bibinfo {volume} {1409}},\ \bibinfo {pages}
  {059} (\bibinfo {year} {2014})},\ \Eprint {http://arxiv.org/abs/1407.3677}
  {arXiv:1407.3677 [hep-ph]} \BibitemShut {NoStop}%
%%CITATION = ARXIV:1407.3677;%%
\bibitem [{\citenamefont {Cyrol}\ \emph {et~al.}(2015)\citenamefont {Cyrol},
  \citenamefont {Huber},\ and\ \citenamefont {von Smekal}}]{Cyrol:2014kca}%
  \BibitemOpen
  \bibfield  {author} {\bibinfo {author} {\bibfnamefont {A.~K.}\ \bibnamefont
  {Cyrol}}, \bibinfo {author} {\bibfnamefont {M.~Q.}\ \bibnamefont {Huber}}, \
  and\ \bibinfo {author} {\bibfnamefont {L.}~\bibnamefont {von Smekal}},\
  }\href {\doibase 10.1140/epjc/s10052-015-3312-1} {\bibfield  {journal}
  {\bibinfo  {journal} {Eur. Phys. J.}\ }\textbf {\bibinfo {volume} {C75}},\
  \bibinfo {pages} {102} (\bibinfo {year} {2015})},\ \Eprint
  {http://arxiv.org/abs/1408.5409} {arXiv:1408.5409 [hep-ph]} \BibitemShut
  {NoStop}%
%%CITATION = ARXIV:1408.5409;%%
\bibitem [{\citenamefont {Gracey}(2017)}]{Gracey:2017yfi}%
  \BibitemOpen
  \bibfield  {author} {\bibinfo {author} {\bibfnamefont {J.~A.}\ \bibnamefont
  {Gracey}},\ }\href {\doibase 10.1103/PhysRevD.95.065013} {\bibfield
  {journal} {\bibinfo  {journal} {Phys. Rev.}\ }\textbf {\bibinfo {volume}
  {D95}},\ \bibinfo {pages} {065013} (\bibinfo {year} {2017})},\ \Eprint
  {http://arxiv.org/abs/1703.01094} {arXiv:1703.01094 [hep-ph]} \BibitemShut
  {NoStop}%
%%CITATION = ARXIV:1703.01094;%%
\bibitem [{\citenamefont {Binosi}\ and\ \citenamefont
  {Quadri}(2012)}]{Binosi:2012st}%
  \BibitemOpen
  \bibfield  {author} {\bibinfo {author} {\bibfnamefont {D.}~\bibnamefont
  {Binosi}}\ and\ \bibinfo {author} {\bibfnamefont {A.}~\bibnamefont
  {Quadri}},\ }\href {\doibase 10.1103/PhysRevD.85.121702} {\bibfield
  {journal} {\bibinfo  {journal} {Phys. Rev.}\ }\textbf {\bibinfo {volume}
  {D85}},\ \bibinfo {pages} {121702} (\bibinfo {year} {2012})},\ \Eprint
  {http://arxiv.org/abs/1203.6637} {arXiv:1203.6637 [hep-th]} \BibitemShut
  {NoStop}%
%%CITATION = ARXIV:1203.6637;%%
\bibitem [{\citenamefont {Cucchieri}\ and\ \citenamefont
  {Mendes}(2012)}]{Cucchieri:2012ii}%
  \BibitemOpen
  \bibfield  {author} {\bibinfo {author} {\bibfnamefont {A.}~\bibnamefont
  {Cucchieri}}\ and\ \bibinfo {author} {\bibfnamefont {T.}~\bibnamefont
  {Mendes}},\ }\href {\doibase 10.1103/PhysRevD.86.071503} {\bibfield
  {journal} {\bibinfo  {journal} {Phys.~Rev.~}\ }\textbf {\bibinfo {volume}
  {D86}},\ \bibinfo {pages} {071503} (\bibinfo {year} {2012})},\ \Eprint
  {http://arxiv.org/abs/1204.0216} {arXiv:1204.0216 [hep-lat]} \BibitemShut
  {NoStop}%
%%CITATION = ARXIV:1204.0216;%%
\bibitem [{\citenamefont {Binosi}\ and\ \citenamefont
  {Theussl}(2004)}]{Binosi:2003yf}%
  \BibitemOpen
  \bibfield  {author} {\bibinfo {author} {\bibfnamefont {D.}~\bibnamefont
  {Binosi}}\ and\ \bibinfo {author} {\bibfnamefont {L.}~\bibnamefont
  {Theussl}},\ }\href@noop {} {\bibfield  {journal} {\bibinfo  {journal}
  {Comput. Phys. Commun.}\ }\textbf {\bibinfo {volume} {161}},\ \bibinfo
  {pages} {76} (\bibinfo {year} {2004})},\ \Eprint
  {http://arxiv.org/abs/hep-ph/0309015} {hep-ph/0309015} \BibitemShut {NoStop}%
%%CITATION = HEP-PH/0309015;%%
\bibitem [{\citenamefont {Binosi}\ \emph {et~al.}(2009)\citenamefont {Binosi},
  \citenamefont {Collins}, \citenamefont {Kaufhold},\ and\ \citenamefont
  {Theussl}}]{Binosi:2008ig}%
  \BibitemOpen
  \bibfield  {author} {\bibinfo {author} {\bibfnamefont {D.}~\bibnamefont
  {Binosi}}, \bibinfo {author} {\bibfnamefont {J.}~\bibnamefont {Collins}},
  \bibinfo {author} {\bibfnamefont {C.}~\bibnamefont {Kaufhold}}, \ and\
  \bibinfo {author} {\bibfnamefont {L.}~\bibnamefont {Theussl}},\ }\href
  {\doibase 10.1016/j.cpc.2009.02.020} {\bibfield  {journal} {\bibinfo
  {journal} {Comput. Phys. Commun.}\ }\textbf {\bibinfo {volume} {180}},\
  \bibinfo {pages} {1709} (\bibinfo {year} {2009})},\ \Eprint
  {http://arxiv.org/abs/0811.4113} {arXiv:0811.4113 [hep-ph]} \BibitemShut
  {NoStop}%
%%CITATION = ARXIV:0811.4113;%%
\end{thebibliography}

\end{document}